\documentclass[twocolumn,showpacs,preprintnumbers,amsmath,amssymb]{revtex4}
\usepackage{graphicx}% Include figure files
\usepackage{dcolumn}% Align table columns on decimal point
\usepackage{bm,color}% bold math

\newcommand{\Tr}[1]{\underset{#1}{\mbox{Tr}}}
\newcommand{\Eref}[1]{Eq.~(\ref{#1})}
\newcommand{\Fref}[1]{Fig.~\ref{#1}}
\newcommand{\figwidth}{3.35in}
\newcommand{\nin}{\in \hspace{-.37em}/ }
\newcommand{\ninn}{\in \hspace{-.90em}/ }
\newcommand{\simleq}{\hspace{0.3em}\raisebox{0.4ex}{$<$}\hspace{-0.75em}\raisebox{-.7ex}{$\sim$}\hspace{0.3em}}

\newcommand{\figcaption}[1]{\def\@captype{figure}\caption{#1}}
\newcommand{\tblcaption}[1]{\def\@captype{table}\caption{#1}}

\begin{document}

\preprint{APS/123-QED}

\title{A statistical-mechanical study of evolution of robustness in noisy environment}% \\¤Ç²þ¹Ô

\author{Ayaka Sakata}
\email{ayaka@huku.c.u-tokyo.ac.jp}
%\altaffiliation[Also at ]{}%Lines break automatically or can be forced with \\
\affiliation{Graduate school of Arts and Sciences, The University of Tokyo, Komaba, Meguro-ku, Tokyo 153-8902, Japan.}%This line break forced with \textbackslash\textbackslash
\author{Koji Hukushima}%
\email{hukusima@phys.c.u-tokyo.ac.jp}
\affiliation{Graduate school of Arts and Sciences, The University of Tokyo, Komaba, Meguro-ku, Tokyo 153-8902, Japan.}%This line break forced with \textbackslash\textbackslash
\author{Kunihiko Kaneko}
\email{kaneko@complex.c.u-tokyo.ac.jp}
%\homepage{http://www.Second.institution.edu/~Charlie.Author}
\affiliation{
Graduate school of Arts and Sciences, The University of Tokyo, Komaba, Meguro-ku, Tokyo 153-8902, Japan.\\
Complex Systems Biology Project, Exploratory Research for Advanced
Technology (ERATO), Japan Science and Technology Agency (JST), Tokyo, Japan}%

\date{\today}% It is always \today, today,
             %  but any date may be explicitly specified

\begin{abstract}
In biological systems, expression dynamics that can provide fitted
phenotype patterns with respect to a specific function have evolved
through mutations. This has been observed in the evolution of proteins
for realizing folding dynamics through which a target structure is
shaped. We study this evolutionary process by introducing a statistical-mechanical
model of interacting spins, where a configuration of spins and their
interactions $\bm{J}$ represent a phenotype and genotype, respectively.
The phenotype dynamics are given by a stochastic process with temperature
$T_{S}$ under a Hamiltonian with $\bm{J}$. The evolution of $\bm{J}$ is also stochastic with
temperature $T_{J}$ and follows mutations introduced into $\bm{J}$
and selection based on a fitness defined for a configuration of a
given set of target spins. Below a certain temperature $T_{S}^{c2}$,
the interactions $\bm{J}$ that achieve the target pattern evolve,
whereas another phase transition is observed at $T_{S}^{c1}<T_{S}^{c2}$.
At low temperatures $T_{S}<T_{S}^{c1}$, the Hamiltonian exhibits
a spin-glass like phase, where the dynamics toward the target pattern
require long time steps, and the fitness often decreases drastically
as a result of a single mutation to $\bm{J}$. In the intermediate-temperature
region, the dynamics to shape the target pattern proceed rapidly and
are robust to mutations of $\bm{J}$. The interactions in this region
have no frustration around the target pattern and results in funnel-type
dynamics. We propose that the ubiquity of funnel-type dynamics, as
observed in protein folding, is a consequence of evolution subjected
to thermal noise beyond a certain level; this also leads to mutational
robustness of the fitness.
\end{abstract}

% PACS, the Physics and Astronomy
                             % Classification Scheme.
%\keywords{Suggested keywords}%Use showkeys class option if keyword
                              %display desired
\maketitle

\section{Introduction}%\label{sec:level1}First-level heading:\protect\\ The line break was forced \lowercase{via} \textbackslash\textbackslash

The function of a biological unit is generally determined by a phenotype,
which is the result of a dynamical process that yields a specific
pattern or structure. The dynamics themselves are governed by a genetic
sequence. In evolution governed by a fixed fitness condition, a phenotype
that gives a higher value for the fitness function is selected. The
genes that produce such a phenotype are transferred to the next generation,
and thus, specific genetic sequences are selected. However, we note,
that the dynamical process producing a phenotype is subject to noise.
Accordingly, the genotype-phenotype mapping is generally stochastic.

For example, genetic information determines the amino acid sequence
for a protein, while the tertiary structure responsible for its functions
is generated only by folding dynamics. By the folding process, a structure
is formed in the protein, and this structure serves as the basis for
the function. The genotype-phenotype mapping is formed by this folding
process, and this mapping is stochastic because of the thermal noise
in the folding process\cite{SY}. Related folding dynamics also occur
in t-RNA, where the influence of thermal noise on genotype-phenotype
mapping has been intensively investigated\cite{Ancel-Fontana}. At
a more macroscopic level, genetic information specifies a gene regulatory
network, which determines the dynamics for the gene expression pattern,
thus giving rise to the phenotype. This gene expression dynamics are
again stochastic because the number of proteins in a cell is not necessarily
very large\cite{geneR}. In fact, the stochasticity for gene expression
of isogenic organisms has been studied extensively\cite{Science,SIYK,KEBC}.
In general, a phenotype that gives rise to some particular function
is generated by a dynamical process that is subject to noise. Hence,
phenotypes of isogenic individual organisms are not necessarily identical,
and therefore, they form a distribution.

Considering that a biological function is generally a result of such
stochastic dynamics, the dynamic process that shapes the function
is expected to be robust under such stochasticity; in other words,
the phenotype for the function will not be sensitive to noise\cite{Alon}.
%Considering a complex dynamics such as a folding process from a heteropolymer with an arbitrary sequence,
%such robustness might not be anticipated, in general. 
However, such robustness is not a general property of dynamics. For
example, the complex folding process of a heteropolymer from an arbitrary
random sequence might not have such robustness. In this sense, the
robustness could be a result of evolution. How can a dynamical process
robust to noise be shaped through evolution?

In addition to being robust to noise, a biological system has to remain
relatively robust to mutations in the genetic sequence that occur
through evolution; the phenotype has to be rather insensitive to changes
in the genetic sequence. Are these two types of robustness correlated?
Does noise in the dynamic process affect the evolution of mutational
robustness? Indeed, a possible relationship between robustness to
noise and robustness to mutation has been discussed \cite{Wagner,Wagner-book,Deem,KK-book,KK-PLoS,KK-chaos,KK-Furusawa},
following the pioneering study by Waddington\cite{Waddington} on
the evolution-development relationship, which is referred to as canalization
and genetic assimilation. However, a theoretical understanding of
the evolution of robustness is still insufficient.

Consider a dynamic process for shaping a target phenotype. To have
robustness to noise in such dynamics, it is ideal to adopt dynamics
in which the target phenotype is reached smoothly and globally from
a variety of initial configurations and is maintained thereafter.
In fact, the existence of such global attraction in the protein folding
process was proposed as a consistency principle by Go \cite{Go} and
as ``funnel'' landscape by Onuchic et al. \cite{PF1,Onuchic},
while similar global attraction dynamics have been discovered recently
in gene regulatory networks \cite{FangLi-Ouyang-Tang} and developmental
dynamics \cite{KK-Asashima}. In spite of the ubiquity of such funnel-like
structures for phenotype dynamics, little is understood about how
these structures are shaped by the evolutionary process \cite{SY}.
We also address this question here and show that it is indeed closely
related to the topic of robustness to noise.

In general, it is possible to utilize a complicated model that agrees
well with biological reality in order to answer the above-mentioned
questions, and this will become necessary in the future. However,
at the present level of understanding, in order to understand the
concepts in the evolution of robustness, we choose to investigate
a rather abstract model that can be made tractable in terms of statistical
physics, that is, a system consisting of $N$ Ising spins interacting
globally. Each spin can take be either up or down, and each configuration
of spins corresponds to a phenotype. %Fitness is a function of these configurations. 
The fitness is given by a function of the configuration of some target
spins, and this fitness yields a biological function. An equilibrium
spin configuration is reached by a certain Hamiltonian that is determined
by the interaction between spins. This interaction is given by genes
and can change by mutation. By selection according to the fitness
function, a Hamiltonian that results in a higher fitness is selected.
Indeed, this type of model has been adopted by Saito et al. \cite{SY},
who utilized it in the study of the evolution of protein folding dynamics,
where the spin configuration corresponds to that of residues in a
peptide chain, and the folding dynamics are given by decreasing the
energy in accordance with the Hamiltonian.

Even though the spin model is abstract, it can account for the basic
structures required to study the evolution of genotype and phenotype,
i.e., gene $\rightarrow$ developmental dynamics subject to noise
$\rightarrow$ phenotype $\rightarrow$ fitness. In comparison with
the gene transcription network model utilized in the study of the
evolution of robustness \cite{Wagner,Wagner-book,KK-PLoS,KK-chaos}, the present
spin model is computationally efficient in that Monte Carlo simulations
and the methods developed in statistical mechanics of spin systems can be applied
to answer the above-mentioned general questions on evolution. In fact,
we will analyze the evolution of robustness with respect to such a
statistical-mechanical framework and define a funnel landscape in
terms of frustration, as developed in spin-glass theory \cite{Nishimori,SGB}.

A shorter version of our results has already been published as a letter \cite{SHK_letter},
in which we propose a scenario, based on our numerical simulations, that
the ubiquity of funnel-type dynamics observed in biological systems is a
consequence of evolutional process under noise beyond a certain level.
In this paper, we further study the spin model in particular on the dependence of the result on the
number of the total spins $N$ and the target spins responsible for the fitness
with extensive numerical simulations.
The results suggest that there exists an optimal ratio of the target
to the total spins to achieve evolution of robustness over a wide
range of temperature.
In other words, some degree of redundant spins that do not contribute
to the fitness
is necessary.
We have also carried out
statistical-mechanical calculations of the fitness landscape,
to provide an interpretation of
the relation between the funnel dynamics and robustness to mutations found in
our numerical simulation. These findings will stimulate
further studies on the understanding on the evolution of robustness from
statistical-physics viewpoints.

This paper is organized as follows. In Sec.\ref{model},
we explain the model setup that captures the essential features of
the evolution. In Sec.\ref{results}, the numerical results for the
model are presented. First, we present the dependence of energy and
fitness on the temperature and selection process. Then, we show that
the evolved Hamiltonians are characterized by the frustration in terms
of the statistical physics of spin systems. We classify three phases
on the basis of the robustness of the fitness to noise and mutation,
and we show that a robust system is realized at an intermediate temperature.
The system-size dependence of these results is also discussed.
The origin of the robustness is studied in Sec. \ref{landscape}
by analytically estimating the fitness landscape by using statistical
mechanics. Finally, in Sec.\ref{conclusions}, the conclusions and
prospects for further development are described.

\section{Model setup}
\label{model}

We introduce a statistical-mechanical spin model in which the phenotype
and genotype are represented by configurations of spin variables $S_{i}$
and an interaction matrix $J_{ij}$, respectively, with $i,j=1,\cdots,N$.
The spins $S_{i}$ and $J_{ij}$ can take one of only two values $\pm1$,
and the interaction matrix is assumed to be symmetric, i.e., $J_{ij}=J_{ji}$.
A set of configurations is denoted by $\bm{S}$ for the phenotype
and by $\bm{J}$ for the genotype. The dynamics of the phenotype are
given by a flip-flop update of each spin with an energy function,
which is defined by the Hamiltonian for a given set of genotypes,
\begin{equation}
H(\bm{S}|\bm{J})=-\frac{1}{\sqrt{N}}\sum_{i<j}J_{ij}S_{i}S_{j}.
\label{Hamiltonian}
\end{equation}
We adopt the Glauber dynamics as an update rule, where the $N$ spins
are in contact with their own heat bath at temperature $T_{S}$. The
Glauber dynamics, satisfying the detailed balance conditions, yields
an equilibrium distribution for a given $\bm{J}$: 
\begin{align}
P(\bm{S}|\bm{J},T_{S})=\frac{e^{-\beta_{S}H(\bm{S}\mid\bm{J})}}{Z_{S}(T_{S})},
\end{align}
where $Z_{S}(T_{S})=\Tr{\bm{S}}e^{-\beta_{S}H(\bm{S}\mid\bm{J})}$
and $\beta_{S}=T_{S}^{-1}$. After a relaxation process, the phenotype
$\bm{S}$ follows from the equilibrium distribution, and it is not
determined uniquely from the genotype $\bm{J}$; rather, it is distributed,
except at zero temperature. The phenotype fluctuation is computed
from the Glauber dynamics, and the resulting equilibrium probability
distribution. Thus, the degree of fluctuation is characterized by
the temperature $T_{S}$.

Next, we introduce evolutionary dynamics for the genotype $\bm{J}$.
The genotype is transmitted to the next generation with some variation,
while genotypes that produce a phenotype with higher fitness are selected.
The time scale for genotypic change is generally much larger than
that of the phenotypic dynamics. 
We assume that the two time scales for the phenotypic expression dynamics
and the genotypic evolutionary dynamics are separated, so that the
variables $\bm{S}$ are well equilibrated within the unit time scale
of the slow variable $\bm{J}$. Then, the fitness should be expressed
by a function of the phenotype $\bm{S}$ that is averaged with respect
to the distribution. Here, we define the fitness as 
\begin{equation}
\Psi(\bm{J}|T_{S})=\Big\langle\prod_{i<j\in\bm{t}}\delta(S_{i}-S_{j})\Big\rangle\equiv\Big\langle\psi\Big\rangle,
\label{Fit}
\end{equation}
where $\langle\cdots\rangle$ denotes the expectation value with
respect to the equilibrium probability distribution. The set $\bm{t}$
denotes a subset of $\bm{S}$ with size $t$ ; the members of $\bm{t}$
are termed as target spins. We refer to configurations such that all
target spins are aligned in parallel as target configurations, which
are assumed to give a requested appropriate function.
By a gauge transformation on the target spin and the corresponding
elements of $\bm{J},$ a choice of any other form of spin alignment
for the fitness function, instead of the ``ferromagnetic'' configuration,
yields the same result \cite{Nishimori}. 
The fitness can be interpreted as the average frequency
of finding the target configurations in equilibrium for a given $\bm{J}$.
It should be noted that in our model, only the target spins contribute
explicitly to the fitness and the remaining spins have no direct influence
on the fitness and the selection of genes. Hence, the spin configuration
for a given fitness has redundancy.

The genotype dynamics are a result of mutations and selection, i.e.,
changes according to the fitness function following random flip-flops
of genes. Hence, for a genetic dynamics, we once again adopt the Glauber
dynamics by using the fitness instead of the Hamiltonian in the phenotype
dynamics, where the genotype $\bm{J}$ is in contact with a heat bath
whose temperature $T_{J}$ is different from $T_{S}$. In specific,
the dynamics for the genotype are given by a stochastic Markov process
with the following stationary distribution: 
\begin{eqnarray}
P(\bm{J}|T_{S},T_{J})=\frac{e^{\beta_{J}\Psi(\bm{J}|T_{S})}}{Z_{J}(T_{S},T_{J})},
\label{J_bunpu}
\end{eqnarray}
where $Z_{J}(T_{S},T_{J})=\Tr{\bm{J}}e^{\beta_{J}\Psi(\bm{J}|T_{S})}$
and $\beta_{J}=T_{J}^{-1}$. According to the dynamics, genotypes
are selected rather uniformly at high values of the temperature $T_{J}$,
irrespective of the fitness, whereas at low values of $T_{J}$, the
genotypes with higher fitness values are preferentially selected.
In this sense, the temperature $T_{J}$ represents the selection pressure
among mutated genotypes. 
\begin{figure}
\includegraphics[angle=270,width=\figwidth]{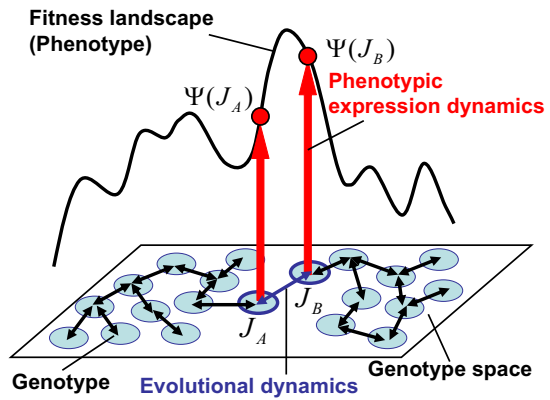} 
\caption{(color online) A schematic representation of our model. The plane
represents the genotype space, where the circles correspond to each
configuration of $\bm{J}$. Their fitness is determined through the
phenotypic expression dynamics, which are given by Glauber-type dynamics.
The landscape in which the height of each point corresponds to the
fitness value is called a fitness landscape. At each generation of
genotype evolutional dynamics, genotypes providing higher fitness
values are selected under the selection pressure $T_{J}$. 
}
\label{geno-pheno} 
\end{figure}

Note that the Glauber dynamics for the genotype $\bm{J}$ is applied
over a much longer time scale than the dynamics for the phenotype
$\bm{S}$; the genotype $\bm{J}$ changes only during the reproduction
of each individual, while the spin dynamics proceed within a developmental
time scale to shape the phenotype. Hence, we update $J_{ij}$ after
the spins are updated a sufficient number of times for attaining the
equilibrium configurations. In actual simulations, a candidate $\bm{J}^{\prime}$
for the next generation is set by some flips of a randomly chosen
$J_{ij}$ from the current $\bm{J}$, while the transition probability
from $\bm{J}$ to $\bm{J}^{\prime}$ is given by Metropolis rules,
$\mbox{min}(1,\exp(\beta_{J}(\Psi(\bm{J})-\Psi(\bm{J}^{\prime})))$.
\Fref{geno-pheno} shows a schematic explanation of our model.

Our model provides two landscapes: the free energy landscape of spins
and the fitness landscape of $\bm{J}$s. The free energy landscape
is determined by a configuration of $\bm{J}$ and the phenotypic expression
dynamics correspond to the relaxation process on the landscape. The
fitness of $\bm{J}$, $\Psi(\bm{J})$, is given by the phenotypic
expression dynamics on the free energy landscape of spins. We call
the landscape of fitness the fitness landscape. The evolutional dynamics
of $\bm{J}$ correspond to a random walk to the top of the fitness
landscape; the random walk is generated by noise whose intensity is
given by $T_{J}$.

A statistical-mechanical spin model similar to ours has been studied
for protein evolution \cite{SY}, where a genetic algorithm is used
for genetic dynamics. Our model, which is based on two equilibrium
distributions, enables us to conduct this study by using a Markov-chain
type simulation, for which a population-based simulation developed
by a genetic algorithm is also available. Further, analytical tools
developed in statistical mechanics are helpful for gaining a better
understanding of the model.

\section{Results}
\label{results}
\subsection{Fitness and Energy}

\begin{figure}
\begin{centering}
\includegraphics[width=\figwidth]{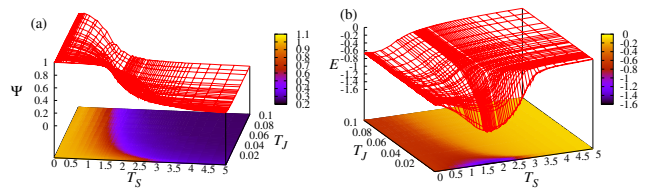}
\par\end{centering}
\caption{(color online) The density plots of the fitness $\Psi(\bm{J})$ and
the energy (both are given as \Eref{eqn:Psi}) are shown in (a)
and (b), respectively, in the $T_{S}-T_{J}$ plane with $N=15$ and
$t=3$. 
}

\label{Fit_contour} 
\end{figure}

We have carried out MC simulations of the model discussed above and
studied the dependence of the fitness and energy on $T_{S}$ and $T_{J}$.
They are given by 
\begin{equation}
\Psi(T_{S},T_{J})=[\Psi(\bm{J}|T_{S})]_{J},\ \ E(T_{S},T_{J})=[\langle H(\bm{S}|\bm{J})\rangle]_{J},
\label{eqn:Psi}
\end{equation}
respectively, where $[\cdots]_{J}$ denotes the average with respect
to the equilibrium probability distribution, $P(\bm{J},T_{S},T_{J})$.
MC sampling with temperature $T_{S}$ under the Hamiltonian $H$ and
the stochastic selection process governed by the fitness are carried
out alternately. In our simulations of the spin dynamics, the exchange
Monte Carlo simulation (EMC) \cite{EMC} is introduced to accelerate
the relaxation time to equilibrium and obtain the equilibrium spin
distribution efficiently. In this section, we concentrate on the analysis
of the equilibrium state. 
\Fref{Fit_contour} (a) and (b) show the dependence of the fitness
and the energy on $T_{S}$ and $T_{J}$, respectively, for $N=15$
and $t=3$.
For each generation of the genotype dynamics, the fitness and energy
are averaged with respect to the equilibrium distribution over 1500
MC steps after discarding the first 1500 MC steps; this number of
steps is sufficient for equilibration. The data are averaged over
the last 1000 generations. The dependence on the system and target
size will be discussed later. For any $T_{S}$, the fitness decreases
monotonically with $T_{J}$, but the rate of decrease is affected
significantly by $T_{S}$. The fitness for sufficiently low $T_{S}$
remains at a high level and decreases only slightly with an increase
in $T_{J}$, while for a medium value of $T_{S}$, the fitness gradually
decreases to a lower level as a function of $T_{J}$, and eventually,
for a sufficiently high value of $T_{S}$, it never reaches a high
level. This result implies that the structure of the fitness landscape
depends on $T_{S}$, the temperature at which the system has evolved.

The energy function, on the other hand, shows a significant dependence
on $T_{S}$. While the energy is represented by a monotonically increasing
function of $T_{S}$, for high $T_{J}$, it exhibits non-monotonic
behavior for low $T_{J}$ and has a minimum at $T_{S}\simeq2.0$.
The configurations that include the target pattern at the energy minimum
are obtained around $T_{S}\simeq2.0$. This non-monotonicity of the
energy corresponds to a negative specific heat in the sense of standard
thermodynamics. This is not possible in
quenched spin systems with fixed $\bm{J}$. However,
the interactions $\bm{J}$ depend on the temperature $T_{S}$ and
$T_{J}$. It would be convenient to obtain an explicit formula for
the derivative of the energy with respect to $T_{S}$:
\begin{equation}
\frac{dE(T_{S},n)}{dT_{S}}=\beta_{S}^{2}\Big\{[\sigma_{E}^{2}]_{J}+\beta_{J}\mbox{Cov}_{\bm{J}}\Big(\langle H\rangle,\mbox{Cov}_{\bm{S}}(\psi,H)\Big)\Big\},\label{hinetsu}\end{equation}
 where $\mbox{Cov}_{\bm{J}}(A,B)=[AB]_{J}-[A]_{J}[B]_{J}$, $\mbox{Cov}_{\bm{S}}(A,B)=\langle AB\rangle-\langle A\rangle\langle B\rangle$
and $\sigma_{E}^{2}=\langle H^{2}\rangle-\langle H\rangle^{2}$. The
first term of \Eref{hinetsu} is the usual specific heat of the
random system and it must be positive, and $T_{S}$-dependence of
$\bm{J}$ comes from the second term, which can be negative. 

The configurations of $\bm{J}$ giving rise to the highest fitness
value generally have a huge redundancy. Using a fluctuation induced
by $T_{S}$, a specific subset of the configurations of $\bm{J}$
with lower energy is selected among the redundant configurations at
around $T_{S}\simeq2.0$.

\subsection{Frustration}

\begin{figure}
\begin{centering}
\includegraphics[width=3.35in]{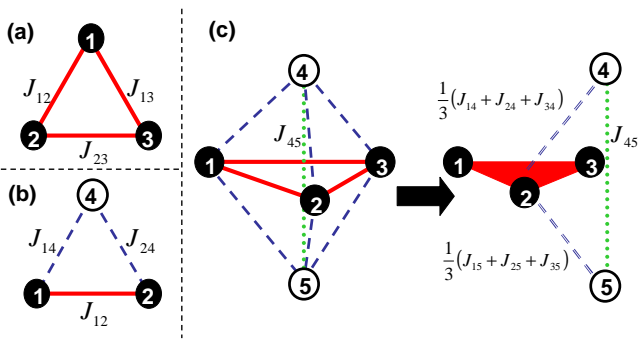} 
\par\end{centering}
\caption{(color online) Definition of frustration in terms of the parameters
$\Phi_{1},\Phi_{2},$ and $\Phi_{3}$ for the case with three target
spins ($t=3$); $S_{i}\ (i=1,2,3)$ are target spins and the remaining
spins are non-targets. The bold lines depict the interactions between
target spins ($\in\bm{J}_{tt}$), the dashed lines depict those between
target and non-target spins ($\in\bm{J}_{to}$), and the dotted lines
depict those between non-target spins ($\in\bm{J}_{oo}$); (a) a triangle
consists of three $\bm{J}_{tt}$ interactions; (b) a triangle consists
of one $\bm{J}_{tt}$ interaction and two $\bm{J}_{to}$ interactions
which are connected to the same non-target spin; (c) a hexahedron
that consists of all types of interactions. By summing up the target
spins, it is represented as a triangle with a renormalized interaction.}
\label{frust} 
\end{figure}

In the medium-temperature range, both a lower energy and a higher
fitness are achieved. What is the structure in $\bm{J}$ configurations
that helps to achieve this? The statistical physics of spin systems tells
us that a decrease in energy implies a decrease in the frustration
in spin configurations. By the definition of the Hamiltonian \Eref{Hamiltonian},
the possible minimum energy is $-C_{2}^{N}\slash\sqrt{N}$, where
$C_{2}^{N}$ is the number of the spin pairs. However, if the interaction
among the three spins satisfies $J_{ij}J_{jk}J_{ki}<0$, the energy
per spin cannot be minimized to the minimum value $-C_{2}^{N}\slash\sqrt{N}$.
Such interactions are said to have frustration\cite{Nishimori,SGB}.
Meanwhile, all the interactions satisfying $J_{ij}J_{jk}J_{ki}>0$
do not have frustration, and the energy of the spin states attains
the minimum value. 
However, the energetically favorable spin configuration cannot be
uniquely determined only by the condition $J_{ij}J_{jk}J_{ki}>0$.
The spin configurations that have low energy should be the target
configurations when both the decrease in energy and the increase in
fitness are simultaneously achieved. 
In our case, the target spins play a distinct role, and therefore,
we need to quantify the frustration while distinguishing between target
and non-target spins; this is in contrast to the standard spin-glass
study.

The interactions are divided into three categories: those between
target spins, $\bm{J}_{tt}\ (\{J_{ij}\mid i,j\in\bm{t}\})$, those
between target and non-target spins, $\bm{J}_{to}\ (\{J_{ij}\mid i\in\bm{t},j\ninn\ \bm{t}\})$,
and those between non-target spins, $\bm{J}_{oo}\ (\{J_{ij}\mid i,j\ninn\ \bm{t}\})$.
It can be assumed that the frustration of all categories decreases
at intermediate $T_{S}$. To confirm this, we should define the conditional
frustration for each category of spins. \Fref{frust}(a) shows the
minimal configuration consisting of the interactions in $\bm{J}_{tt},$
(b) shows that consisting of the interactions in $\bm{J}_{tt}$ and
$\bm{J}_{to}$, and (c) shows that consisting of the interactions
in $\bm{J}_{to}$ and $\bm{J}_{oo}$. 

\begin{figure*}
\begin{centering}
\includegraphics[width=\textwidth]{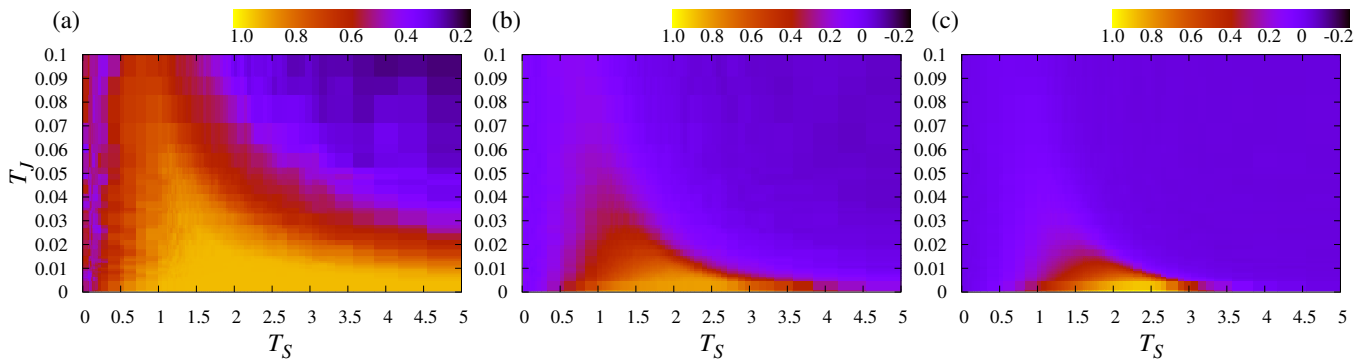} 
\par\end{centering}
\caption{(color online) Density plot of local frustrations: (a) $\Phi_{1}$,
(b) $\Phi_{2},$ and (c) $\Phi_{3}$ in the $T_{S}-T_{J}$ plane.
The data are computed by averaging over 150 genotypes $\bm{J}$ evolved
at given temperatures $T_{S}$ and $T_{J}$ with $N=15$ and $t=3$.}
\label{mu_c} 
\end{figure*}

We first define $\Phi_{1}$ as the frequency of positive coupling
among target spins, i.e., \begin{align}
\Phi_{1}(T_{S},T_{J})=\frac{2}{t(t-1)}\Big[\sum_{i<j\in\bm{t}}J_{ij}\Big]_{J}.\label{J1}\end{align}
 The target configurations are energetically preferred under ferromagnetic
coupling, i.e., $\Phi_{1}=1$, for which no frustration exists among
the target spins (\Fref{frust}(a)).

Second, we define $\Phi_{2}$ as 
\begin{equation}
\Phi_{2}(T_{S},T_{J})=\frac{2}{t(t-1)(N-t)}\Big[\sum_{i<j\in\bm{t}}\sum_{k\nin\bm{t}}J_{ik}J_{jk}\Big]_{J},
\label{J2}
\end{equation}
where $t(t-1)(N-t)/2$ is the total number of possible spins: two
target spins and one non-target spin. Here, $\Phi_{2}=1$ implies
that no frustration exists in the interactions between target and
non-target spins, and thus, the target configuration is at an energy
minimum even when these interactions are included (\Fref{frust}(b)).

Lastly, as a measure of the frustration among non-target spins, $\Phi_{3}$
is defined as 
\begin{equation}
\Phi_{3}(T_{S},T_{J})=\frac{1}{C_{2}^{N-t}}\Big[\sum_{k<l\nin\bm{t}}\Big(\frac{1}{t}\sum_{i\in\bm{t}}J_{ik}\Big)J_{kl}\Big(\frac{1}{t}\sum_{j\in\bm{t}}J_{jl}\Big)\Big]_{J},
\label{J3}
\end{equation}
where $C_{2}^{N-t}$ is the total number of possible pairs of non-target
spins. \Fref{frust}(c) helps to comprehend the definition of $\Phi_{3}$.
Each non-target spin interacts with all the target spins, and it has
$t$ interactions that are categorized into $\bm{J}_{to}$, e.g.,
$J_{14},J_{24}$ and $J_{34}$. By summing up all the interactions,
the frustration is computed as $\frac{1}{9}(\sum_{i=1}^{3}J_{i4})J_{45}(\sum_{j=1}^{3}J_{5j})$.
By considering all the possible non-target spins instead of sites
4 and 5, $\Phi_{3}$ is defined as \Eref{J3}. If $\Phi_{3}$ is
equal to 1, the frustration is not introduced by the interactions
between non-target spins; in other words, there is no frustration
globally. Hence, the system with $\Phi_{1}=\Phi_{2}=\Phi_{3}=1$ is
in the Mattis state \cite{Mattis}, which can be transformed to ferromagnetic
interaction by a gauge transformation.

For the interaction $\bm{J}$, with evolution under an environment
with temperature $T_{S}$, we have computed $\Phi_{1},\Phi_{2}$,
and $\Phi_{3}$ by performing MC simulations. In \Fref{mu_c}, we
present contour maps of (a) $\Phi_{1}(T_{S},T_{J})$, (b) $\Phi_{2}(T_{S},T_{J})$,
and (c) $\Phi_{3}(T_{S},T_{J})$ in the $T_{S}-T_{J}$ plane. 
At sufficiently low
$T_{J}$, the frustration parameters attain the maximum value 1 at
the intermediate $T_{S}$, where the frustrations are extensively
eliminated, while they remain finite at low $T_{S}$ and high $T_{S}$.
We define the intermediate temperature region as $T_{S}^{c1}<T_{S}<T_{S}^{c2},$
where the frustration parameter $\Phi_{2}$ equals 1.
These temperatures depend on $T_{J}$, and we express them as $T_{S}^{c1}(T_{J})$
and $T_{S}^{c2}(T_{J})$. We see that for low $T_{J}$ ($\simleq0.05$),
the phase diagram is split into three phases. The first one is frustrated
and adapted phases for $T_{S}<T_{S}^{c1}(T_{J})$. For $T_{S}<T_{S}^{c1}(T_{J})$,
all $\Phi_{i}~(i=1,2,3)$ are less than unity, and hence, the frustration
remains for target and non-target spins.

\begin{figure}
\begin{centering}
\includegraphics[width=\figwidth]{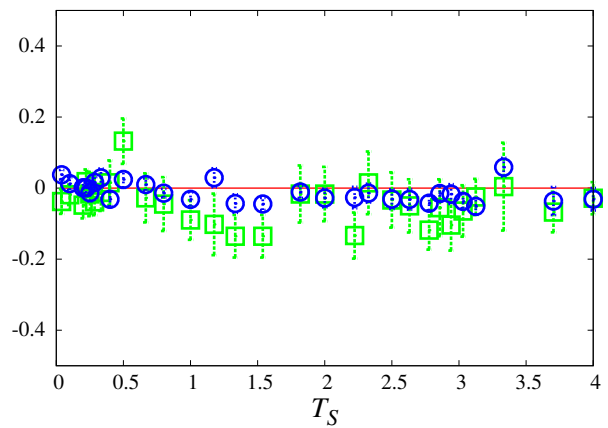} 
\par\end{centering}
\caption{(color online) $T_{S}$-dependence of the averages $\overline{\bm{J}_{to}}$
$(\square)$ and $\overline{\bm{J}_{oo}}$ $(\bigcirc)$ for a fixed
$T_{J}=0.5\times10^{-3}$. 
}
\label{pattern_zero} 
\end{figure}

For $T_{S}\geq T_{S}^{c1}(T_{J})$, $\Phi_{1}$ equals 1, so that
a target configuration is embedded as an energetically favorable state
(\Fref{mu_c}(a)). For a finite system with finite $T_{J}$, $\Phi_{j}$
cannot be exactly 1. However, as long as $T_{J}$ is low, the deviation
of $\Phi_{j}$ from 1 at the intermediate temperature is negligible.
In contrast to $\Phi_{1}$, the sum of the $J_{ij}$ in $\bm{J}_{to}$
and $\bm{J}_{oo}$ fluctuates around $0$ at any $T_{S}$. \Fref{pattern_zero}
shows the averages $\overline{\bm{J}_{to}}=[\sum_{i\in\bm{t},j\in\bm{o}}J_{ij}]_{J}$
and $\overline{\bm{J}_{oo}}=[\sum_{i\in\bm{o},j\in\bm{o}}J_{ij}]_{J}$
of the summation of $J_{ij}$ in $\bm{J}_{to}$ and $\bm{J}_{oo}$,
respectively, at a low $T_{J}$ ($0.5\times10^{-3}$). As shown in
\Fref{pattern_zero}, $\overline{\bm{J}_{to}}$ and $\overline{\bm{J}_{oo}}$
do not deviate from 0 at any $T_{S}$. This implies that no specific
patterns are embedded in the spin configuration apart from the target
spins.

\begin{figure}
\begin{centering}
\includegraphics[width=\figwidth]{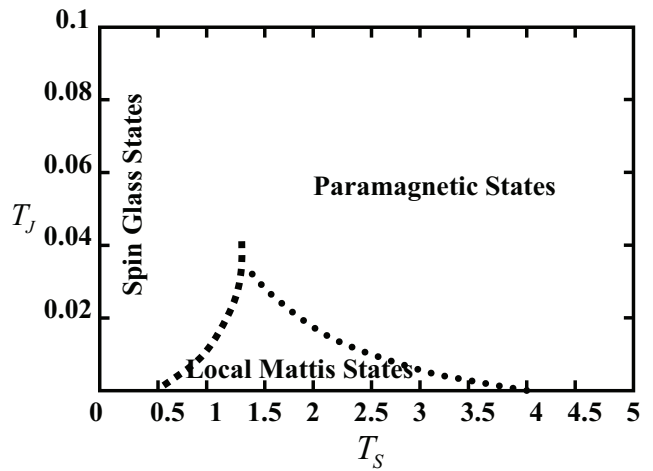} 
\par\end{centering}
\caption{Phase diagram of the evolved $\bm{J}$s at $N=15$ and $t=3$. Three types of evolved $\bm{J}$
are defined on the basis of the value of the fitness and $\Phi_{2}$.
Their properties are summarized below in Table \ref{table:phase}.}
\label{figure:phase} 
\end{figure}

For $T_{S}^{c1}(T_{J})\leq T_{S}\leq T_{S}^{c2}(T_{J})$, $\Phi_{2}$
is also equal to 1, implying that the frustration among spins is not
introduced via interactions with a non-target spin (\Fref{mu_c}(b)).
In this temperature range, $\Phi_{3}$ is not always equal to 1, except
for $T_{S}\sim2.0$, where the Mattis state arises (\Fref{mu_c}(c)).
When $\Phi_{2}=1$ and $\Phi_{3}\neq1$, the frustration is not completely
eliminated from the non-target spin interactions $\bm{J}_{oo}$; this
is in contrast to the Mattis state. We call such a $\bm{J}$ configuration
``local Mattis state,'' and it is characterized by $\Phi_{1}=\Phi_{2}=1$
but $\Phi_{3}\neq1$. This implies that the interactions $\bm{J}$
have no frustration around the target spins, but there is some frustration
between non-target spin interactions. The interactions $\bm{J}$ required
to form such a local Mattis state are obtained as a consequence of
the evolution around $T_{S}^{c1}(T_{J})\leq T_{S}\leq T_{S}^{c2}(T_{J})$
for low $T_{J}$, where both the fitted target configuration and lower
energy are achieved. 
The $T_{S}$ range in which the local Mattis state is stabilized becomes
narrower with an increase in $T_{J}$.
The phase diagram of the model is shown in Fig.~\ref{figure:phase}. 

For $T_{S}>T_{S}^{c2}(T_{S})$, the frustration parameter $\Phi_{2}$
is less than 1, and consequently, the frustration remains, and the
fitness $\Psi$ starts to decrease and the energy increases. Thus,
neither adaptation nor energy minimization is achieved.
The parameters $\Phi_{i}$s should converge to 0 as $T_{S}\to\infty$,
because for random $\bm{J}$, the numbers of frustrated and non-frustrated
loops are equal. In fact, at $T_{S}\sim5.0$, $\Phi_{1}$ also starts
to decrease.

\subsection{Relaxation dynamics}
\label{relax-dynamics}

\begin{figure}
\includegraphics[width=\figwidth]{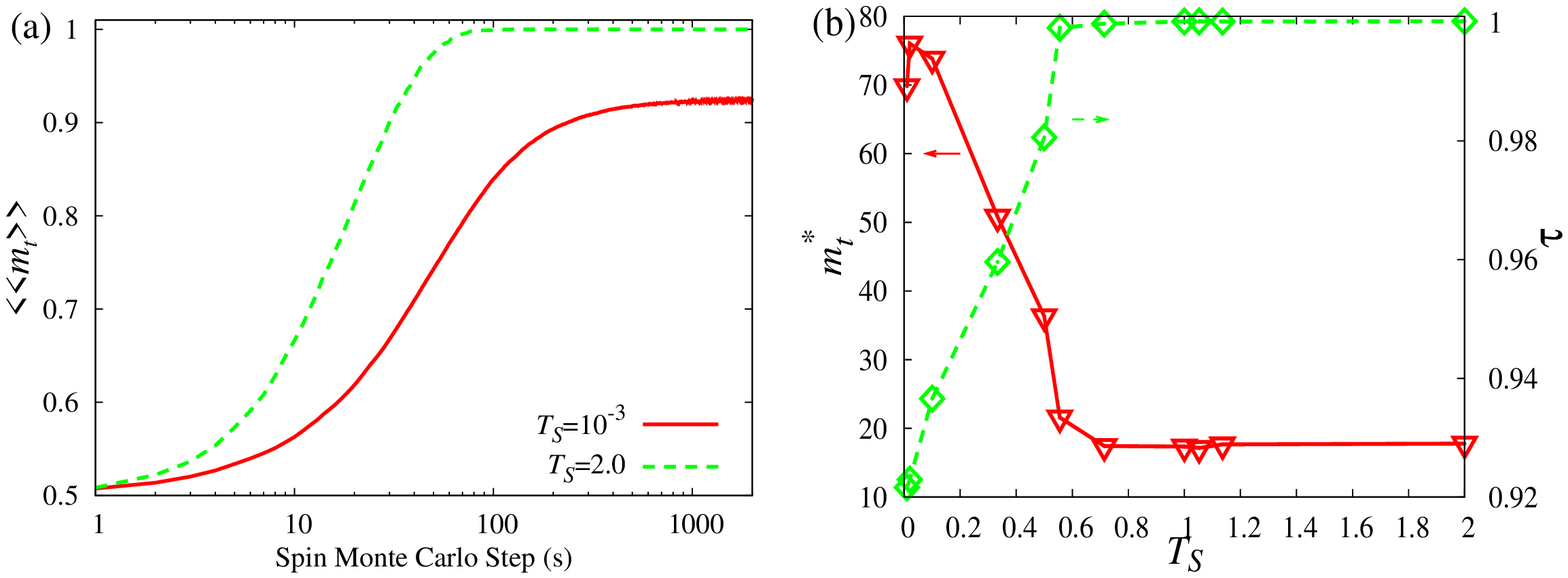} 
\caption{(color online) (a) Relaxation dynamics of the averaged magnetization
of target spins, $\langle\langle m_{t}\rangle\rangle,$ averaged over
the adapted interactions $\bm{J}$ for $T_{S}=10^{-3}$ (solid curve) and $T_{S}=2.0$ (dashed curve).
The magnetization $\langle\langle m_{t}\rangle\rangle$ is evaluated
by calculating the average over 30 initial conditions for each $\bm{J}$
and 1000 different samples of $\bm{J}$ that are drawn from $P(\bm{J},T_{S},T_{J}=10^{-3})$.
\protect \\
 (b)The $T_{S}$-dependence of the estimated convergent value of
$\langle\langle m_{t}\rangle\rangle$, $m_{t}^{\ast}$ on the right
axis and the relaxation time $\tau$ on the left. The relaxation time
is estimated from the time constant of an exponential decay of $\langle\langle m_t \rangle\rangle$.}
\label{Fit_kanwa} 
\end{figure}

Thus far, we have computed the fitness in the equilibrium state by
using EMC for accelerating the relaxation dynamics of spins. Under
standard Glauber dynamics, the process may require much longer time
steps. Here, we discuss the relaxation dynamics of spins for adapted
interactions $\bm{J}$ that are obtained from evolution under the
condition of given $T_{S}$ and $T_{J}$. 
The target magnetization $m_{t}=|\frac{1}{t}\sum_{i\in\bm{t}}S_{i}|$
is computed as a function of time. Note that we do not use EMC here;
rather, we adopt standard MC to directly observe the energy landscape
of $\bm{J}$ adapted through evolution. We calculate the average of
$m_{t}$ over $\bm{J}$ drawn from an equilibrium distribution $P(\bm{J},T_{S},T_{J})$
at $T_{S}$ and $T_{J}$. 
\Fref{Fit_kanwa} (a) shows the relaxation dynamics of $\langle\langle m_{t}\rangle\rangle$
for $T_{S}=10^{-3}(\leq T_{S}^{c1})$ and $T_{S}=2.0(T_{S}^{c1}\leq T_{S}\leq T_{S}^{c2})$,
where $\langle\langle\cdots\rangle\rangle$
denotes the average over the initial conditions randomly chosen and
over interactions $\bm{J},$ according to $P(\bm{J},T_{S},T_{J})$.
In the simulation, we choose a sufficient low $T_{J}(=10^{-3})$ so
that the obtained interactions have high fitness values. A common
working temperature $T_{S}^{\prime}(=10^{-5})$ for relaxation is
also chosen to be very low in order to examine the $T_{S}$-dependence
of the adapted interactions $\bm{J}$ at $T_{S}$ and $T_{J}$. 
By performing this simulation, the landscape properties of the typical
$\bm{J}$ adapted at $T_{S}$ are clearly determined from a dynamical
viewpoint. 
As shown in \Fref{Fit_kanwa}, the relaxation process of $\langle\langle m_{t}\rangle\rangle$
for low $T_{S}$ is much slower even when the working temperatures
$T_{S}'$ are the same. Furthermore, the relaxation process 
converges to a value $m_{t}^{\ast}$ that is less than 1 and remains
at that value for a long time. The deviation of $\langle\langle m_{t}\rangle\rangle$
from 1 gives the fraction of the initial conditions that fails to
reach the target within this time span, because each $m_{t}$ for
$t=3$ is either 1 or $1\slash3$ depending on whether the target
configuration is reached. Indeed, the relaxation dynamics are strongly
dependent on the initial conditions. For some initial conditions,
the spins are trapped at a local minimum, and as a result, the target
configuration is not realized over a long time span. After a much
longer time span, $\langle\langle m_{t}\rangle\rangle$ approaches
1, the equilibrium value, when the spins are updated under the temperature
$T_{S}$, i.e., the temperature adopted for evolution. Such dependence
on initial conditions is not observed for $\langle\langle m_{t}\rangle\rangle$
when $T_{S}>T_{S}^{c1}$, where $\langle\langle m_{t}\rangle\rangle$
approaches 1 rather quickly.

From an estimate of the convergent value of the target magnetization,
$m_{t}^{\ast},$ within a given time scale, we obtain the relaxation
time $\tau$ by fitting the estimates to the function $\langle\langle m_{t}\rangle\rangle(s)=m_{t}^{\ast}+c\exp(-s\slash\tau)$,
where $s$ is the Monte Carlo step of the spin dynamics. The parameters
$m_{t}^{\ast}$ and $\tau$ are plotted against $T_{S}$ in \Fref{Fit_kanwa}(b),
which shows that $\tau$ starts to increase and $m_{t}^{\ast}$ decreases
from 1 as $T_{S}$ decreases below $T_{S}^{c1}$. These results imply
that the interactions $\bm{J}$ whose energy landscapes are rugged,
similar to the energy landscape of a spin-glass phase, are dominant
for $T_{S}\leq T_{S}^{c1}$, whereas those with a smooth landscape
around the target are dominant for $T_{S}^{c1}\leq T_{S}\leq T_{S}^{c2}$.
The latter can be interpreted as a type of funnel landscape. Our result
supports the occurrence of transitions from the spin-glass phase to
the funnel phase at $T_{S}^{c1}$caused by thermal fluctuation. Note
that the evolutional formation of funnel from rugged landscapes was
also observed by Saito et al. in the evolution simulations of spin
systems for protein folding \cite{SY}.

\subsection{Robustness to mutation}
\label{Robust-mutation}

We now examine the mutational robustness of the evolved genotypes
in detail. The robustness to mutations corresponds to the stability
of the fitness of $\bm{J}$ with respect to changes in the $\bm{J}$
configuration. From the genotypes $\bm{J}$ that are generated by
$P(\bm{J}|T_{S},T_{J})$, mutations are imposed by flipping the sign
of a certain fraction of randomly chosen matrix elements in $\bm{J}$.
The value of the fraction corresponds to the mutation rate $\mu$.
We evaluate the fitness of the mutated $\bm{J}^{\prime}(\bm{J},\mu)$
at $T_{S}^{\prime}$, i.e., 
\begin{align}
[\Psi(\bm{J}^{\prime} & (\bm{J},\mu)|T_{S}^{\prime})]_{J(T_{S},T_{J})}\nonumber \\
 & =\Tr{\bm{J}}P(\bm{J}|T_{S},T_{J})\frac{\Tr{\bm{S}}\psi e^{-\beta_{S}^{\prime}H(\bm{S}|\bm{J}^{\prime}(\bm{J},\mu))}}{Z(T_{S}^{\prime},\bm{J}^{\prime}(\bm{J},\mu))},
\label{fit_fake}
\end{align}
 where $\beta_{S}^{\prime}=1\slash T_{S}^{\prime}$ and $Z(T_{S}^{\prime},\bm{J}^{\prime})=\Tr{\bm{S}}e^{(-\beta_{S}^{\prime}H(\bm{S}|\bm{J}^{\prime}))}$.
The bracket $[\cdots]_{J(T_{S},T_{J})}$ is almost identical to that
denoted by $[\cdots]_{J}$ defined above; however, the additional
subscript $(T_{S},T_{J})$ indicates the temperatures at which the
genotype $\bm{J}$ evolves. 
If $\mu=0$ and $T_{S}^{\prime}=T_{S}$, $[\cdots]_{J(T_{S},T_{J})}$
is equal to the usual fitness defined in \Eref{eqn:Psi}; $[\Psi(\bm{J}^{\prime}(\mu=0)|T_{S}^{\prime}=T_{S})]_{J(T_{S},T_{J})}=\Psi(T_{S},T_{J})$.
In order to distinguish the mutational robustness from thermal noise,
we set $T_{J}=0.5\times10^{-3}$ to ensure that the fitness value
with $\mu=0$ is 1 and the working temperature is $T_{S}^{\prime}=10^{-5}$.
The fitness averaged over 150 samples of mutated $\bm{J}^{\prime}$
is plotted against the mutation rate $\mu$ in Fig. 4 in \cite{SHK_letter}
for $T_{S}=10^{-4}$ and $T_{S}=2.0$. 
For low $T_{S}$, the fitness of mutated $\bm{J}$ exhibits a rapid
decrease with an increase in the mutation rate, but when $T_{S}$
is between $T_{S}^{c1}$ and $T_{S}^{c2}$, the fitness does not decrease
until the mutation rate reaches a specific value. We define $\mu_{c}(T_{S})$
as a threshold mutation rate beyond which the fitness $[\Psi(\bm{J}^{\prime}(\bm{J},\mu)|T_{S}^{\prime})]_{J(T_{S},T_{J})}$
drops below 1. 
The value $\mu_{c}$ has a plateau at $T_{S}^{c1}\leq T_{S}\leq T_{S}^{c2}$.
The range of temperatures that result in mutational robustness as
evolution proceeds agrees with the range giving rise to the local
Mattis state, where $\Phi_{2}$ is unity. In other words, mutational
robustness is realized for a set of genotypes with no frustration
around the target spins. The evolution to a mutationally robust genotype
$\bm{J}$ is possible only when the phenotype dynamics are subjected
to noise in the range $T_{S}^{c1}\leq T_{S}\leq T_{S}^{c2}$.

\begin{table}
\begin{tabular}{|c|c|c||c|c|}
\hline 
Phase  & Adaptation  & Frustration  & Landscape  & Robustness \tabularnewline
\hline
\hline 
SG  & Adapted  & Frustrated  & Rugged  & Not robust \tabularnewline
\hline 
LMS  & Adapted  & Not frustrated  & Funnel  & Robust \tabularnewline
\hline 
PM  & Not adapted  & Frustrated  & Rugged  & Not robust \tabularnewline
\hline
\end{tabular}

\caption{Three types of $\bm{J}$---spin-glass (SG), local Mattis state (LMS),
and paramagnetic state (PM)---are defined by the value of the local
frustration parameter $\Phi_{2}$ (\Fref{figure:phase}). The adaptation
and frustration of $\bm{J}$s have been studied in the previous sections,
and their landscape and robustness at a working temperature $T_{S}^{\prime}$
have been studied in this section. We have adopted a very low temperature
$T_{S}^{\prime}$ to reveal the energy and fitness landscape given
by the interaction matrix $\bm{J}$ evolved at each temperature $T_{S}$.
The $\bm{J}$s evolved at $T_{S}<T_{S}^{c1}$ (these $\bm{J}$s belong
to the SG phase) have high fitness values and frustrations. Their
adaptation is not robust to noise and mutation because of their rugged
landscapes. The $\bm{J}$s evolved at $T_{S}^{c1}<T_{S}<T_{S}^{c2}$,
(these {$\bm{J}$}s belong to the LMS phase) have high fitness values
and less frustrations. They are robust to noise and mutation, and
their landscapes give funnel-type dynamics for spins. 
The fitness values of the $\bm{J}$s evolved at $T_{S}>T_{S}^{c2}$
(these $\bm{J}$s belong to the PM phase) cannot be high, and the $\bm{J}$s
have frustrations. }

\label{table:phase} 
\end{table}

To summarize, there are three phases, local Mattis, spin glass and
paramagnetic disorder ones, for the evolved $\bm{J}$, as shown in
\Fref{figure:phase}.
Each phase defined by the frustration parameter $\Phi_{2}$ has a
distinct characteristic feature in relaxation dynamics and some
robustness, that 
are summarized in Table \ref{table:phase}.

\subsection{Size dependence and the existence of an optimal target size to obtain LMS}

\begin{figure*}
\includegraphics[width=0.32\textwidth]{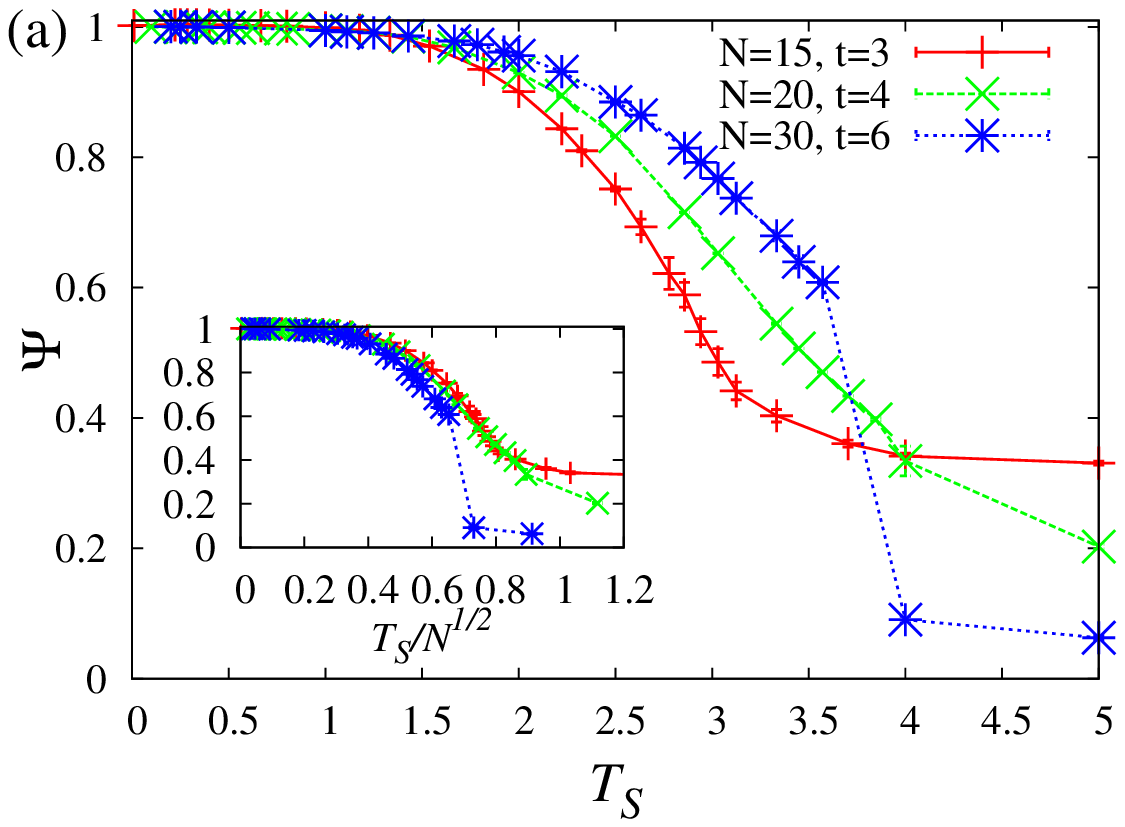}
\includegraphics[width=0.32\textwidth]{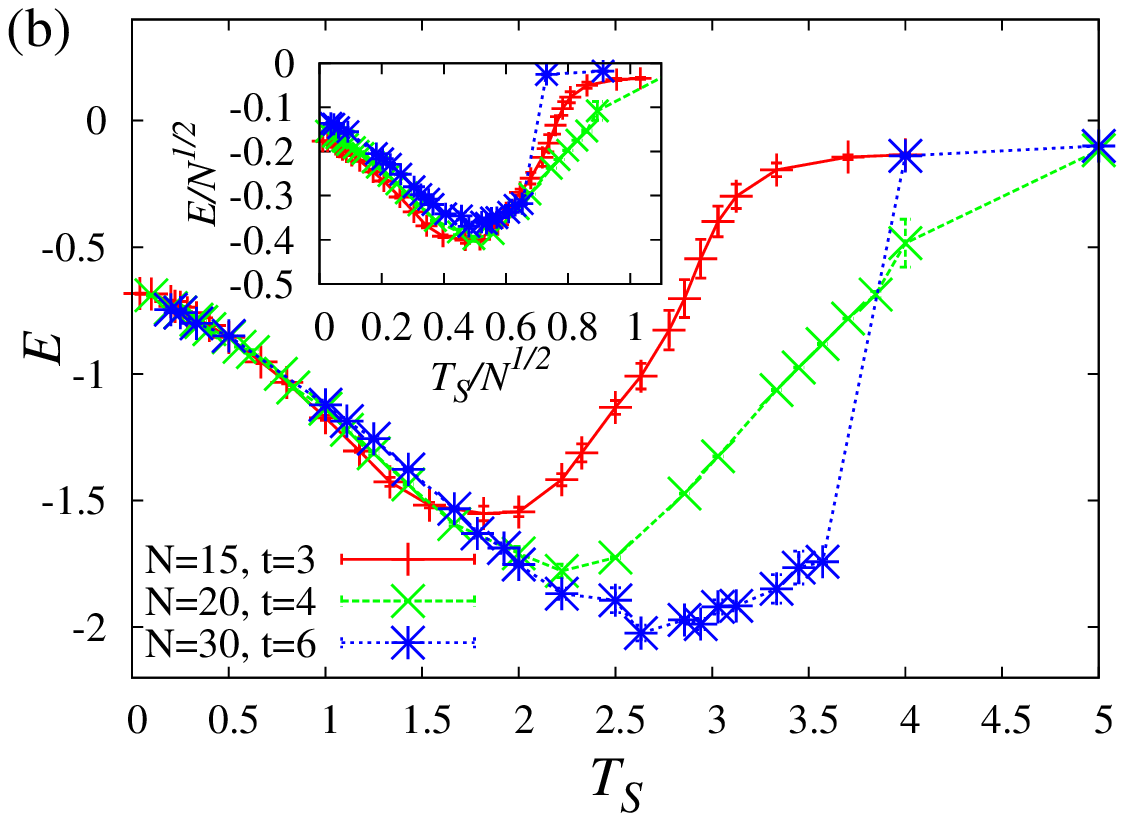}
\includegraphics[width=0.32\textwidth]{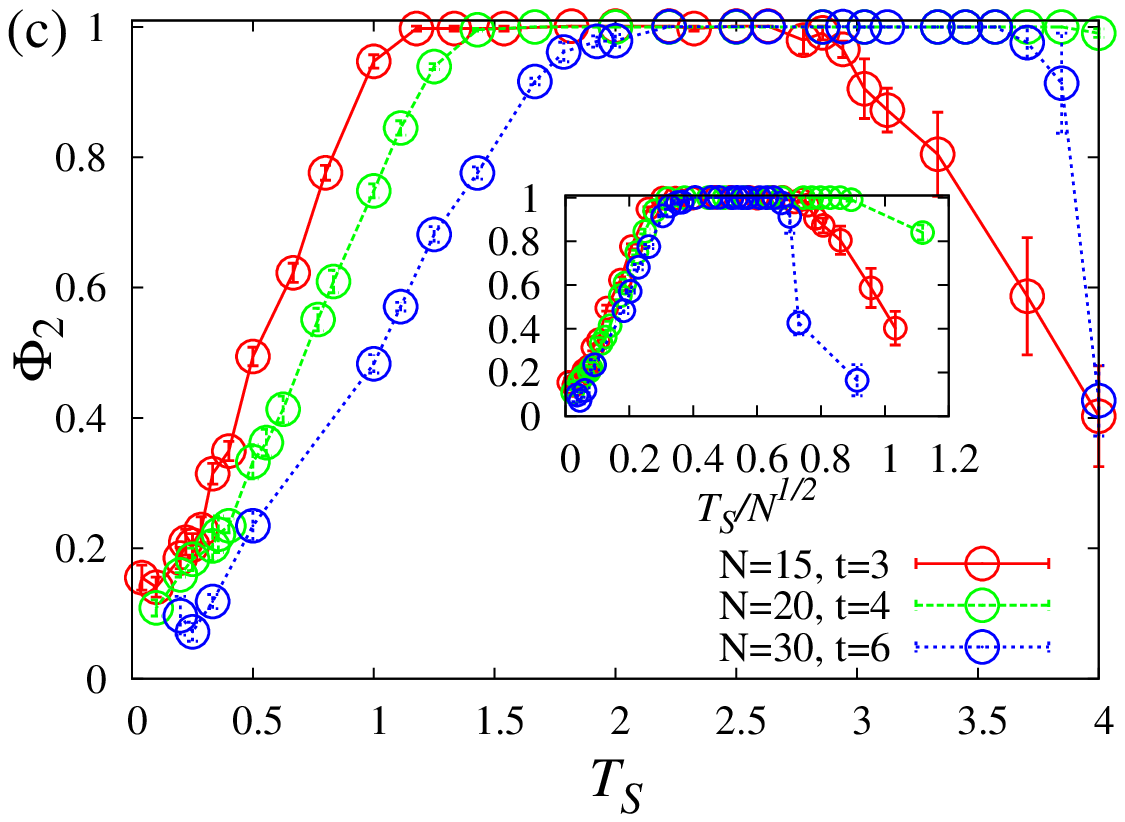} 
\caption{(color online) The $T_{S}$-dependence of (a) the fitness, (b) the
energy, and (c) the frustration parameter $\Phi_{2}$ at $N=15,~t=3$, $N=20,~t=4$
and $N=30,t=6$. The temperature $T_{J}$ is fixed at $0.5\times10^{-3}$.
The inset of each figure shows the dependence of each quantity on
the temperature rescaled by $\sqrt{N}$. The vertical axis of the
inset of (b) is also rescaled by $\sqrt{N}$.}
\label{N30t6} 
\end{figure*}

\begin{figure*}
\includegraphics[width=0.32\textwidth]{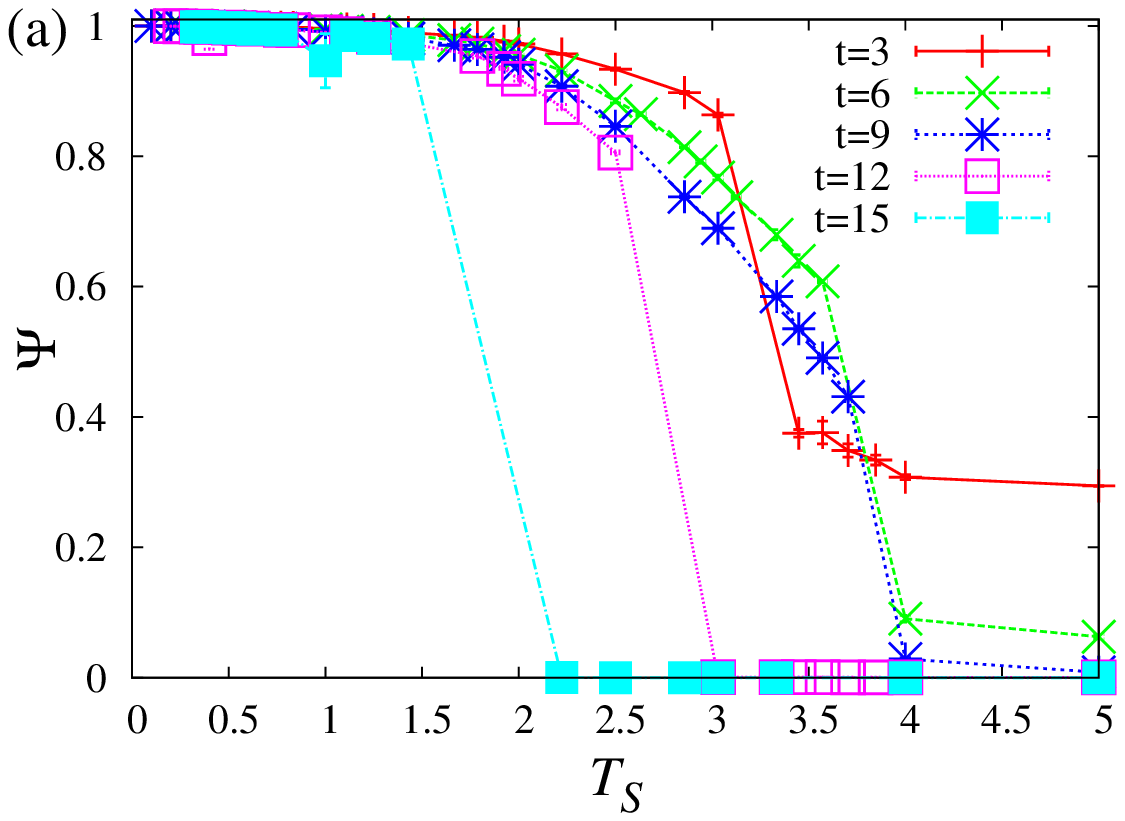}
\includegraphics[width=0.32\textwidth]{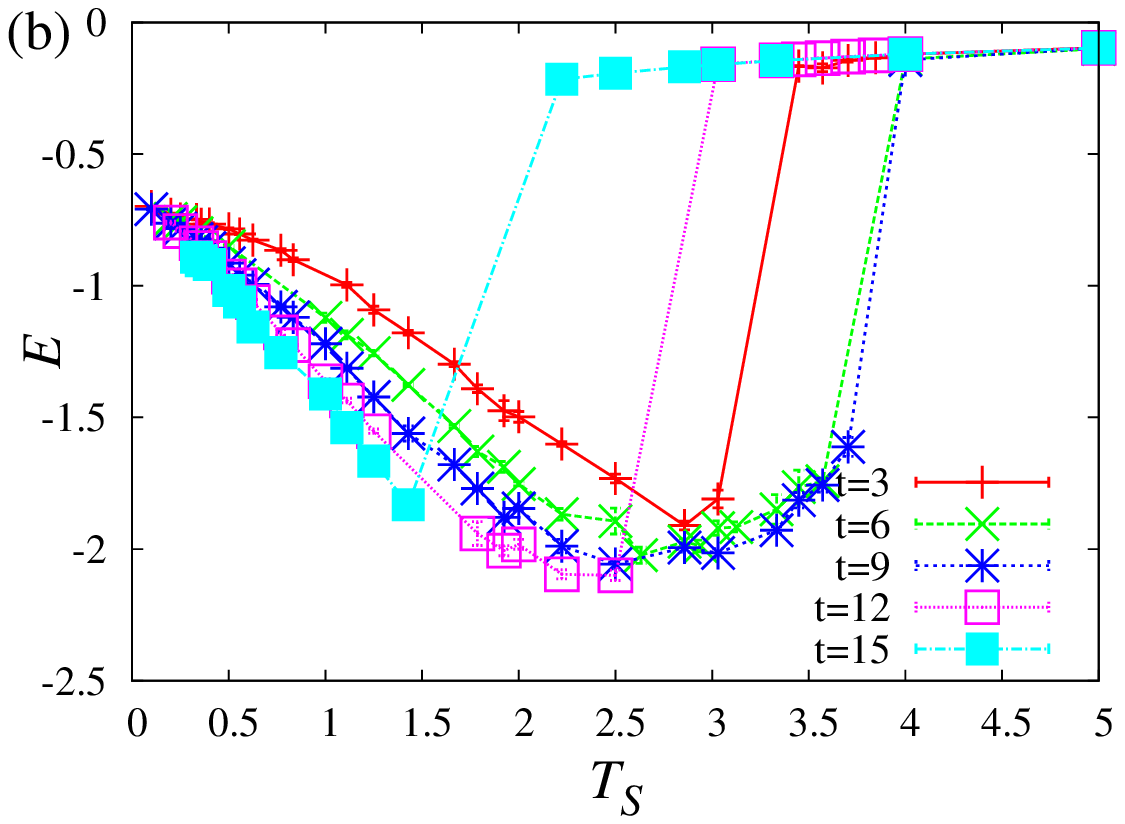}
\includegraphics[width=0.32\textwidth]{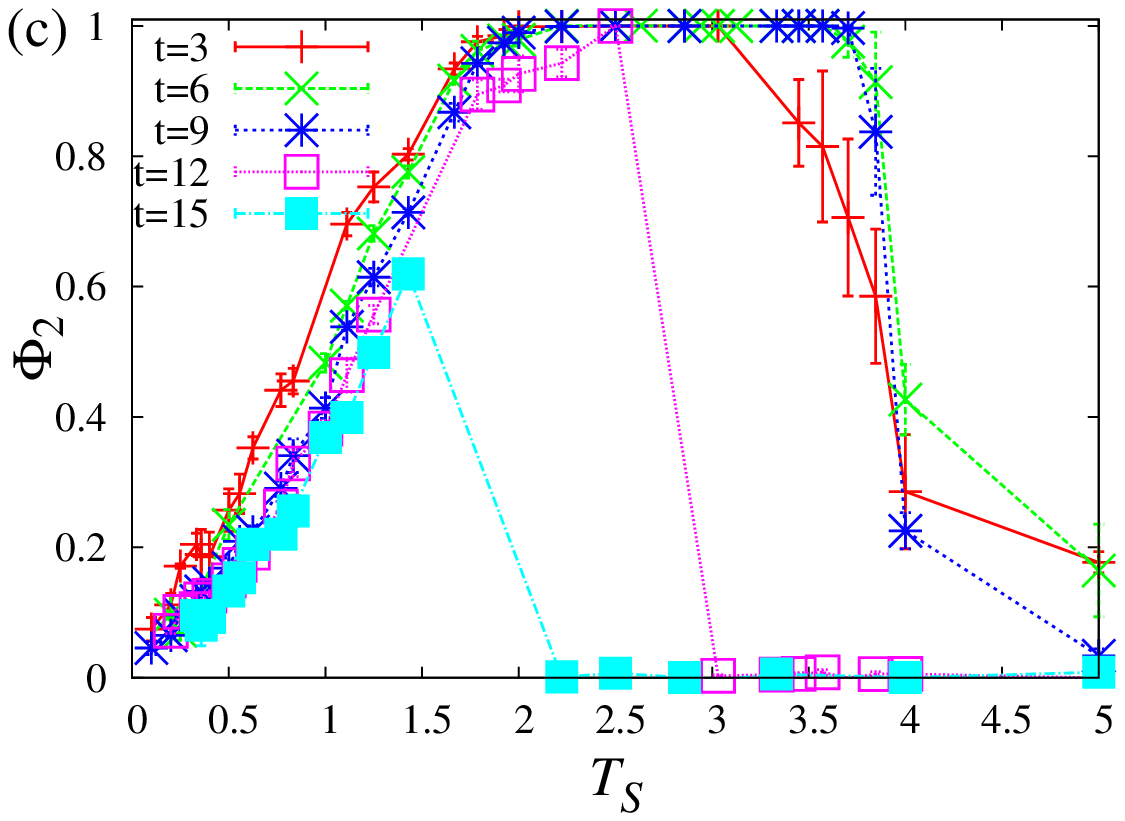} 
\caption{(color online) The $T_{S}$-dependence of (a) the fitness, (b) the
energy, and (c) $\Phi_{2}$ at $N=30$ and $t=3,6,9,12$, and $15$.
The temperature $T_{J}$ is fixed at $0.5\times10^{-3}$. 
The convergence value of the fitness (a) at $T_S\to\infty$ is $2^{-t+1}$.}
\label{N30}
\end{figure*}

To check the generality of the transition to the local Mattis state
as well as the mutational robustness, we examine the model for three
system sizes, $N=15,~20$, and $30$; the ratio $t\slash N=0.2$. In
\Fref{N30t6}, we compare the $T_{S}$-dependence of (a) the fitness,
(b) the energy, and (c) one of the frustration parameters $\Phi_{2}$ at
$N=15~(t=3)$, $N=20~(t=4)$ and $N=30~(t=6)$ at a fixed $T_{J}=0.5\times10^{-3}$.
As shown in \Fref{N30t6}, the behavior of these quantities is qualitatively
similar, and by rescaling the temperature $T_{S}$ by a factor $\sqrt{N}$,
the fitness, energy, and $\Phi_{2}$ lines merge into a single line
until the PM phase appears (insets of \Fref{N30t6}). The plateau
$\Phi_{2}=1$ exists at all system sizes we have studied, and we show
that the rescaled temperature $T_{S}^{c1}(N)\slash\sqrt{N}$ fit together.

The rescaling factor is $\sqrt{N}$ since the order of energy changes
from $O(N)$ to $O(N^{3\slash2})$ by the evolution at the intermediate
$T_{S}$ because of the existence of the local Mattis states. We have
defined the Hamiltonian \Eref{Hamiltonian} with the normalization
coefficient $1\slash\sqrt{N}$; in this definition, we consider the
average over $\bm{J}$ that would be drawn from an i.i.d set of $\bm{J}$s.
However, at intermediate $T_{S}$, the distribution of $\bm{J}$ deviates
from the uniform distribution, and local Mattis states appear with
high probability. As a result, the order of energy changes, and the
temperature $T_{S}^{c1}(N)$ is proportional to $\sqrt{N}$.

\begin{figure}
\begin{center}
\includegraphics[width=\figwidth]{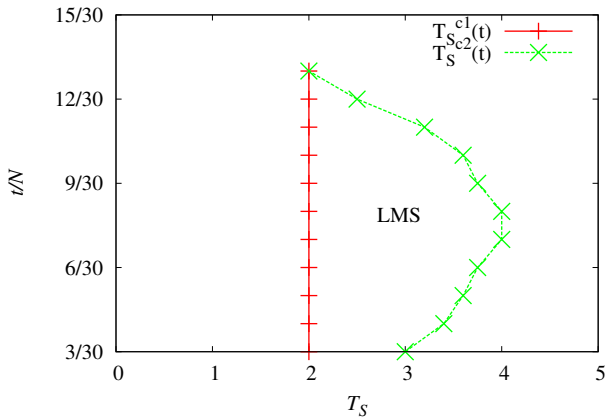}
\end{center}
\caption{(color online) 
$t\slash N - T_S$ diagram of the domain where the local Mattis states are 
found at $N=30$ and a fixed $T_J=0.5\times 10^{-3}$.
The points on the boundary of the domain are estimated by the values of $T_S$
at which $\Phi_2$ drops from 1, as in \Fref{N30}(c),
and the line is guide to eye.
}
\label{t-TS}
\end{figure}

Next, we change the number of target spins $t$ while fixing
$N$ at 30. \Fref{N30} shows (a) the fitness, (b) the energy, and
(c) the frustration parameter $\Phi_{2}$, at $N=30$ and $3\leq t\leq 15$. 
The threshold temperature $T_{S}$ at which the fitness
value decreases rapidly increases as $t$ increases from a small value
to 9 and decreases for larger $t$. The $T_{S}$-dependence of the
energy shows non-monotonic behavior, indicating the existence of the
local Mattis states for all the values of $t$ studied here. 
Similar to the threshold
temperature obtained from the fitness value, the temperature at which
the energy takes a minimum value shows non-monotonic behavior as $t$
increases. Further, the range of the plateau with $\Phi_{2}=1$ is
maximized at $t=7$ and 8. 
Eventually,
the plateau vanishes at $t=12$ at least. The PM phase, where the
adaptation is not acquired, extends toward the lower temperature region
with an increase in $t$; the LMS phase becomes narrower as $t$ increases.
\Fref{N30} shows that the LMS exists only up to $t\simleq0.4N$,
and for $t>0.4N$, a direct transition from the SG to PM phase occurs
with an increase in $T_{S}$. 
The LMS region in which robustness and high
fitness can be achieved is largest at $t\sim N\slash 4$. These findings
indicate that there exists an upper limit of $t$ over which the local
Mattis states cannot be obtained and that there exists a suitable
value of $t$ for stabilizing the local Mattis states for a wide range
of $T_{S}$.
The above simulation results are summarized in
\Fref{t-TS}, where the region of local Mattis state is displayed
in the $t\slash N-T_S$ plane, by fixing $N$ and $T_J$ at
30 and $0.5 \times 10^{-3}$, respectively.

Finally, we remark on the equilibration time of the evolutional dynamics of
$\bm{J}$ to achieve an adapted state, by varying the value of $t$.
At a fixed $T_J$, the time scale increases with the increase of $t$,
even for a fixed size
$N$. Particularly, this increase is significant in the SG phase, while it is
moderate in LMS phase. This might remind us of slow relaxation of the
phenotype dynamics discussed in the previous section.

\section{Fitness landscape}
\label{landscape} We now explain why mutational robustness is realized
only in the intermediate range of temperatures $T_{S}.$ We do so
by performing a statistical-mechanical calculation of the number of
fitted states in order to obtain a sketch of the fitness landscape
for $\bm{J}$. We estimate the degeneracy of the states with highest
fitness as $T_{S}\to0$. Let $W(E_{n}(\bm{J}))$ and $W_{\Psi}(E_{n}(\bm{J}))$
be the number of $n$-th excited states and the number of the target
configurations of $n$-th excited states for a given $\bm{J}$. Then,
the fitness in the limit as $T_{S}\to0$ is given by \begin{align}
\Psi(\bm{J},T_{S}) & =\frac{\sum_{n}W_{\Psi}(E_{n}(\bm{J}))e^{-\beta_{S}E_{n}(\bm{J})}}{\sum_{n}W(E_{n}(\bm{J}))e^{-\beta_{S}E_{n}(\bm{J})}}\equiv\frac{Z_{\Psi}(\bm{J},T_{S})}{Z(\bm{J},T_{S})}\nonumber \\
 & \longrightarrow\frac{W_{\Psi}(E_{0}(\bm{J}))}{W(E_{0}(\bm{J}))}.\label{part-func}\end{align}
Accordingly, as $T_{S}\rightarrow0$, the highest fitness, i.e., unity,
is achieved if and only if $\bm{J}$ satisfies the condition $W_{\Psi}(E_{0}(\bm{J}))=W(E_{0}(\bm{J}))$.
In fact, a large number of $\bm{J}$ satisfy this condition besides
the local Mattis state. For example, let us consider a Mattis state
$\bm{J}$ with no frustration at all and introduce several changes
in the sign of the bonds between target spins $\bm{J}_{tt}$, target
and non-target spins $\bm{J}_{to}$, and non-target spins $\bm{J}_{oo}$.
This procedure, if applied only to the bond flips to $\bm{J}_{oo}$,
produces the local Mattis states. 

\begin{figure}
\begin{centering}
\includegraphics[width=\figwidth]{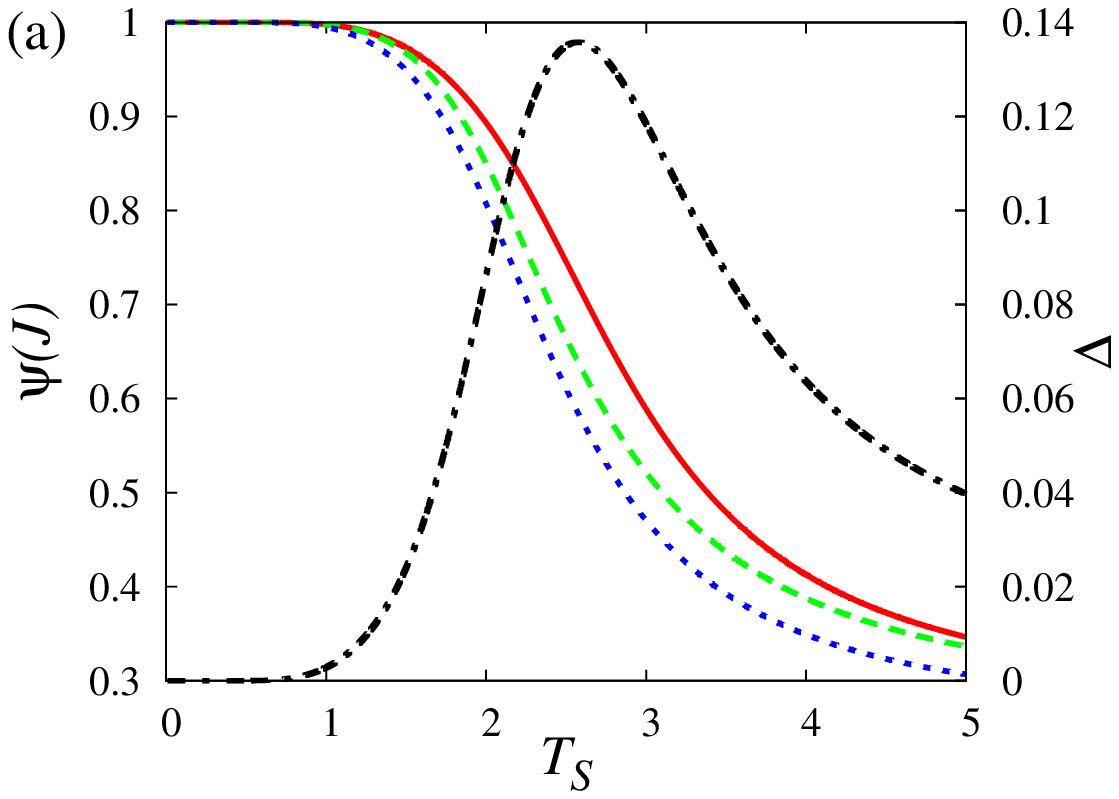}
\includegraphics[width=\figwidth]{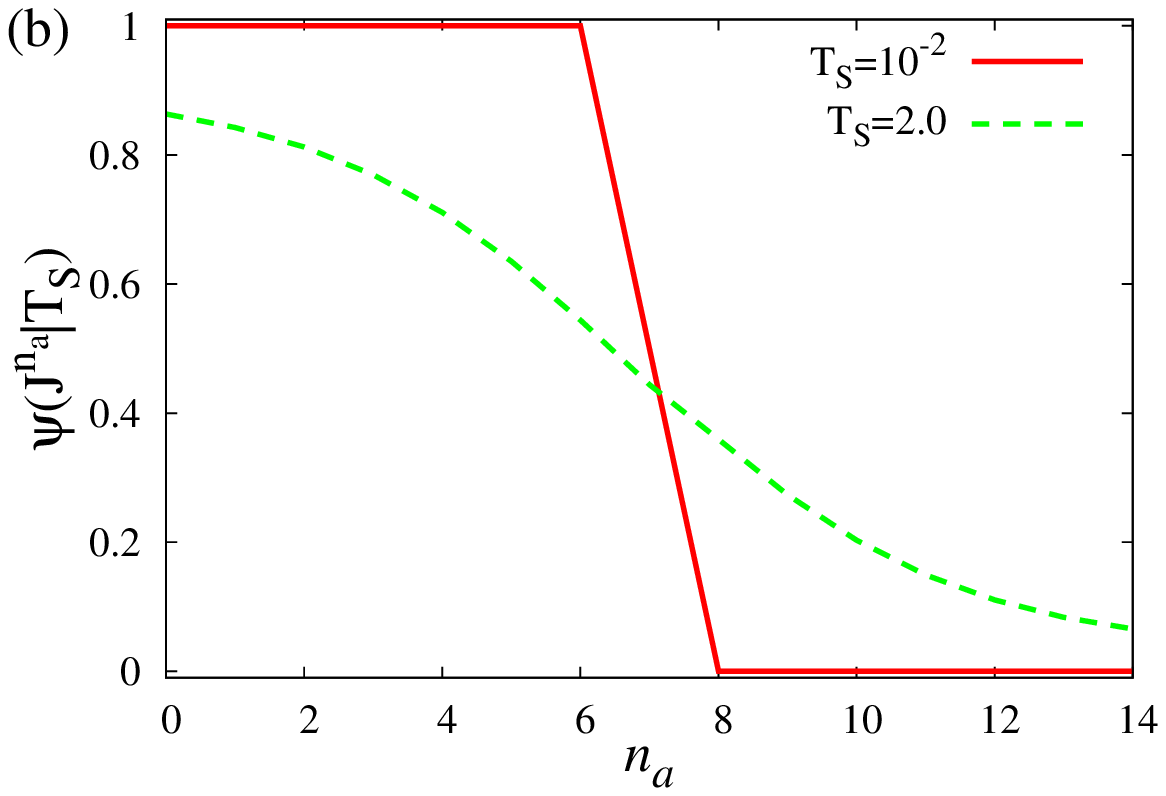} 
\par\end{centering}

\caption{(color online) (a) The $T_{S}$-dependence of the fitness of the Mattis
state (bold line, $\Phi_{1}=\Phi_{2}=\Phi_{3}=1$), one of the local
Mattis states (dashed line, $\Phi_{1}=\Phi_{2}=1,\Phi_{3}=0.89$),
and one of the target-frustrated states (dotted line, $\Phi_{1}=0.333,\Phi_{2}=1,\Phi_{3}=0.89$),
plotted on the left axis. The dashed-dotted line represents the difference
$\Delta$ between the fitness of the Mattis state and that of the
target-frustrated state; $\Delta$ is plotted on the right. \protect \\
 (b) The fitness of $\tilde{\bm{J}}^{n_{a}}$ adapted at $T_{S}=10^{-2}$ (solid line)
and $T_{S}=2.0$ (dashed line) with $N=15$ and $t=3$, plotted as a function
of the number $n_{a}$ of flipped bonds of $\tilde{\bm{J}}^{n_{a}}$.
At low $T_{S}$, the fitness decreases rapidly for $n_{a}=7$ and
decreases gradually at intermediate $T_{S}$. 
}

\label{S3} 
\end{figure}

First, we calculate the fitness of the Mattis states ($\Phi_{1}=\Phi_{2}=\Phi_{3}=1$),
and we consider representative examples of the local Mattis states
($\Phi_{1}=\Phi_{2}=1$,$\Phi_{3}\neq1$) and target-frustrated states
that have frustration among the target spins and hence $\Phi_{1}\neq1$.
Their fitness is given as the ratio of the ``conditioned
partition function'' $Z_{\Psi}(\bm{J})$ and the partition function
$Z(\bm{J})$, as given by \Eref{part-func}. The location of the
frustrations does not influence the partition function, but it influences
$Z_{\Psi}(\bm{J})$. To determine their fitness function, we should
derive the partition functions $Z(\bm{J})$ and $Z_{\Psi}(\bm{J})$
for three types of $\bm{J}$.

At first, the partition function of $\bm{J}$ with $x$ frustrated
interactions, $Z(x,T_{S})$ is given as 
\begin{align}
Z(x,T_{S})=2c(x)\sum_{i=0}^{x}C_{i}^{x}f(i,N),
\label{eqn:part-func}
\end{align}
where 
\begin{align}
f(i,N)=\Big(4e^{\frac{2\beta_{S}}{\sqrt{N}}}\sinh(\frac{2\beta_{S}}{\sqrt{N}})\Big)^{i}\sum_{n=i}^{N-i}e^{-\frac{\beta_{S}}{\sqrt{N}}(N-n)n}C_{n-i}^{N-2i},
\label{Zf_Mattis}
\end{align}
and $c(x)=e^{\beta_{S}(-2x+C_{2}^{N})\slash\sqrt{N}}$. This expression
is valid for the case where there is at most one flipped bond at each
site. Therefore, $x$ should be less than $N\slash2$, and the expression
\Eref{eqn:part-func} is efficient, independent of how to assign
the value $x$ for $\bm{J}_{tt}$, $\bm{J}_{to}$, and $\bm{J}_{oo}$.

Next, the conditioned partition function of the local Mattis state with
$x$ frustrated $\bm{J}_{oo}$ interactions (denoted as $Z_{\Psi}^{LMS}(x,T_{S})$)
is given as 
\begin{align}
Z_{\Psi}^{LMS}(x,T_{S})=2c(x)\sum_{i=0}^{x}C_{i}^{x}f(i,N-t).
\label{eqn:Zf}
\end{align}
This expression is valid for the case where there is at most one
flipped bond at each site. Accordingly, $x$ should be less than
$(N-t)/2$. The fitness of the local Mattis states with $x$ frustrated
interactions in $\bm{J}_{oo}$ is given as $Z_{\Psi}^{LMS}(x,T_{S})\slash Z(x,T_{S})$.
Furthermore, the conditioned partition function of the target frustrated
states that have $y$ frustrations in $\bm{J}_{tt}$ (implying $\Phi_{1}\neq1$)
and $x$ frustrations in $\bm{J}_{oo}$ (denoted as $Z^{TF}(x,y,T_{S})$),
is derived from \Eref{Zf_Mattis}. When a bond between a pair of
target spins is flipped from the Mattis state and frustration is generated
among the target spins, the energy of the target configuration increases
by $2\slash\sqrt{N}$; therefore, \begin{equation}
Z_{\Psi}(x,y,T_{S})=Z_{\Psi}^{LMS}(x,T_{S})\times e^{-\frac{2\beta_{S}}{\sqrt{N}}y},\end{equation}
 where we again consider that each target spin is connected at most
one flipped bond and $y=1,\cdots,t\slash2$. The partition functions
of the target-frustrated states with $x$ frustrated $\bm{J}_{oo}$
bonds and $y$ frustrated $\bm{J}_{tt}$ bonds are given by $Z(x+y,T_{S})$,
and their fitnesses are given by $Z_{\Psi}^{TF}(x,y,T_{S})\slash Z(x+y,T_{S})$.

\Fref{S3} shows the $T_{S}$-dependence of the typical fitness
values for Mattis, local Mattis, and target-frustrated states. The
difference between the fitness of the target-frustrated state and
that of the Mattis state, which is denoted as $\Delta$, is also plotted
as a function of $T_{S}$. The value of fitness always approaches
unity as $T_{S}\to0$, whereas such degeneracy is split by an increase
in $T_{S}$. From the difference in fitness between the Mattis and
the target-frustrated states, $\Delta,$ the ratio of the probabilistic
weight between these states is obtained as $\exp(\beta_{J}\Delta)$.
This suggests that fewer frustrated $\bm{J}$ states around the target,
i.e., the local Mattis states, are preferentially selected only at
the intermediate temperature. 

Next, we introduce frustration into $\bm{J}_{to}$ and denote the
constructed $\bm{J}$ as $\tilde{\bm{J}}_{t}^{n_{a}}$, where the
superscript $n_{a}$ represents the number of altered bonds and the
subscript $t$ represents the condition in which the altered bonds
exist between the target and non-target spins, i.e., $\bm{J}_{to}$.
Therefore, the frustration parameter $\Phi_{2}$ for $\bm{J}_{to}$
of the state $\tilde{J}_{t}^{n_{a}}$ does not equal 1. The state
$\tilde{J}_{t}^{0}$ is simply the original Mattis state, which has
the highest fitness, whereas for $n_{a}=N-1$, direct computation
shows that the fitness is the least. Again, from a straightforward
calculation, it can be shown that the fitness of $\tilde{J}_{t}^{n_{a}}$
remains to be the highest fitness up to $n_{a}\leq N\slash2$. Hence,
there is a region in the $\bm{J}$-state space connected by a single
point mutation (change of sign in a single element in $\bm{J}$) in
which the fitness retains its highest value. We refer to this region
as the neutral space\cite{Ancel-Fontana,Wagner,NE}, in the sense
that a mutation within the region is neutral. Note that in addition
to this construction, there is more degeneracy among the fittest $\bm{J,}$
as shown in \Fref{S3}(a).

We have computed how the fitness decreases as $\bm{J}$
is changed to leave the neutral space, for $T_{S}\rightarrow0$. In
\Fref{S3}(b), the fitness of $\tilde{J}_{t}^{n_{a}}$ is plotted
as a function of the number of altered bonds $n_{a}$. By just a single
point mutation, the fitness decreases suddenly to its lowest value
at some $n_{a}$. This implies the existence of a clear edge in the
neutral space. 
The genotype located at the edge of the neutral space is not robust
to mutation. It is obvious that the Mattis state, i.e., the genotype
located at the center of the neutral space, is robust to mutation.
However, since the fitness of genotypes is constant throughout the
neutral space, both the robust genotypes at the center of the neutral
space and the non-robust genotypes at the edge are selected with equal
weight. Then, there is no selection pressure to eliminate genotypes
that are at the edge of the neutral space. 
Although complete degeneracy holds only for $T_{S}\rightarrow0$,
the above argument is valid for sufficiently low temperatures, and
therefore, the robustness to mutation cannot be expected to exist
in the spin-glass phase at low $T_{S}$.
This is also related to the fact that long time scale for equilibration
in the evolutional dynamics is required in the spin-glass phase. As
shown in \Fref{S3}(b), the fitness value around an adapted $\bm{J}$
decreases abruptly against the mutation.
The number of $\bm{J}$ configurations on the plateau with $\Psi = 1$
that appear at low $T_S$ is roughly estimated as $2^{N-t}$ because of
the gauge transformations for $N-t$ sites, while the total number of
possible $\bm{J}$ configurations is $2^{N^2}$.
The ratio $2^{N-t}\slash 2^{N^2}$ is strongly suppressed as $N$
increases.
This implies that the evolution of $\bm{J}$ hardly finds the plateau by
the local update.

\begin{figure}
\begin{centering}
\includegraphics[width=\figwidth]{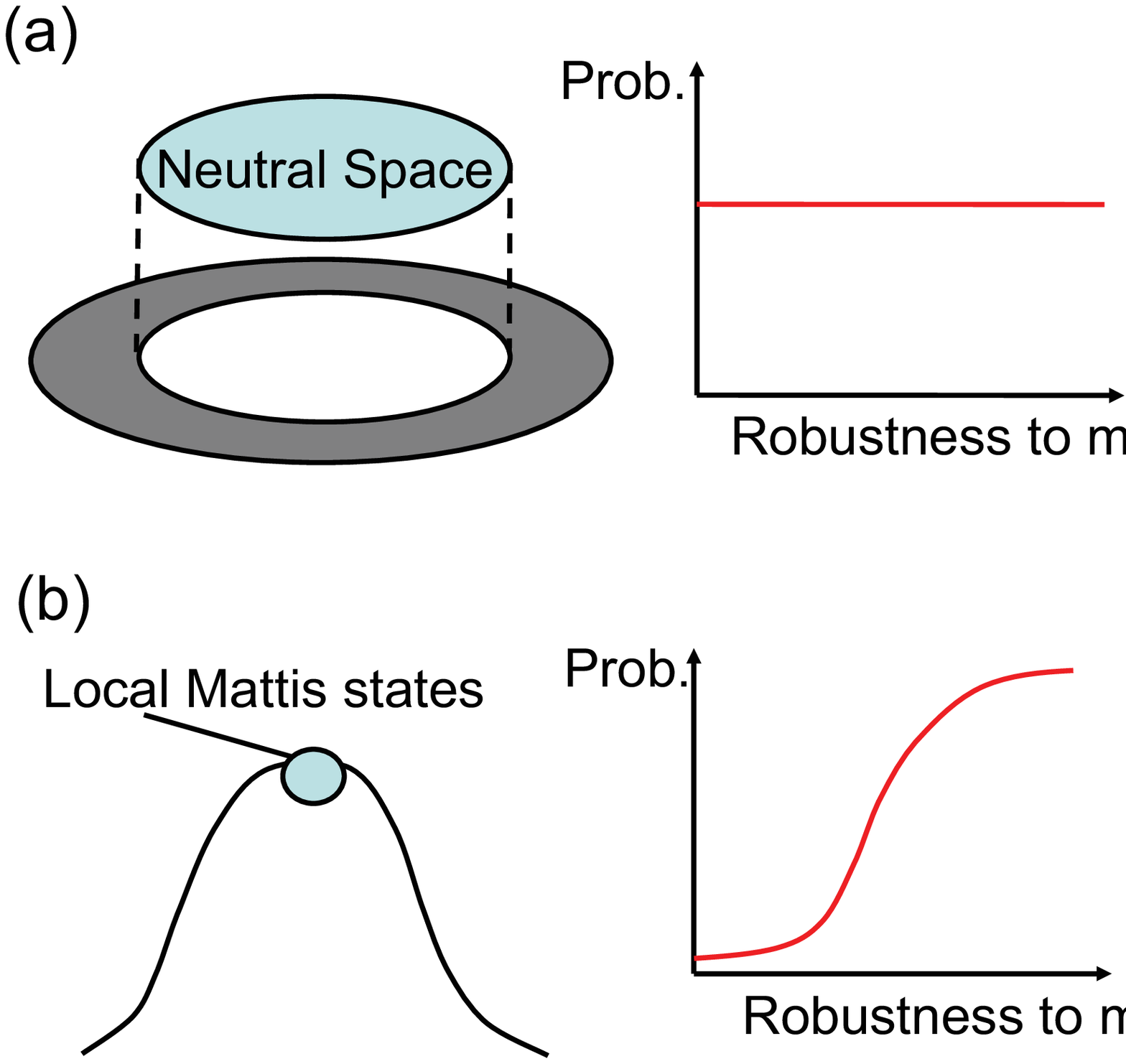} 
\par\end{centering}
\caption{(color online) Schematic representation of the fitness landscape and
the probability distribution of the robustness to mutation estimated
from the fitness landscape. \protect\protect \\
 (a) At low $T_{S}$, the genotypes that satisfy the condition
$W_{\Psi}(E_{0}(\bm{J}))=W(E_{0}(\bm{J}))$ exist in the neutral space,
where the fitness attains its highest value. The local Mattis states
also exist in the neutral space, but they constitute a negligible
fraction compared to the frustrated genotypes. The neutral space is
surrounded by non-adapted genotypes with sharp boundaries, as in \Fref{S3}(b).
The genotypes around the edge of the neutral space that have very
low robustness to mutation survive in an evolutionary sense, similar
to the robust genotypes located at the center of neutral space, that
are much fewer in number. \protect \\
 (b) For $T_{S}$ with $T_{S}^{c1}\leq T_{S}\leq T_{S}^{c2}$,
the degeneracy observed at low values of $T_{S}$ is split, and the
local Mattis states have the highest fitness. The fitness of the local
Mattis state decreases continuously with increasing mutation. There
is a correlation between fitness and robustness, i.e., the local Mattis
state with the highest fitness is the most robust to mutation, and
the robust genotypes are selected preferentially.}
\label{neutral_space} 
\end{figure}

In contrast, at intermediate $T_{S}$, the fitness landscape is not
neutral. For example, the fitness of $\tilde{\bm{J}}_{t}^{n_{a}}$
gradually decreases from its highest value with increasing $n_{a}$,
as shown in \Fref{S3}(b). There is selection pressure toward the
genotype with $n_{a}=0$. The genotypes with larger $n_{a}$ have
both lower robustness to mutation and lower fitness, but less of
such genotypes are selected. Hence, the evolution toward higher fitness
also induces robustness to mutation, as a result of the correlation
between fitness and robustness.

On the basis of the above argument, schematic representations of the
fitness landscape at low and intermediate $T_{S}$, together with
the distribution of mutational robustness, are shown in \Fref{neutral_space}(a)
and (b), respectively. The mutational robustness at an intermediate
temperature observed in MC simulations, as described in the previous
section, is thus interpreted as a consequence of the evolution of
the fitness landscape at such temperatures.

\begin{figure}
\begin{center}
\includegraphics[width=\figwidth]{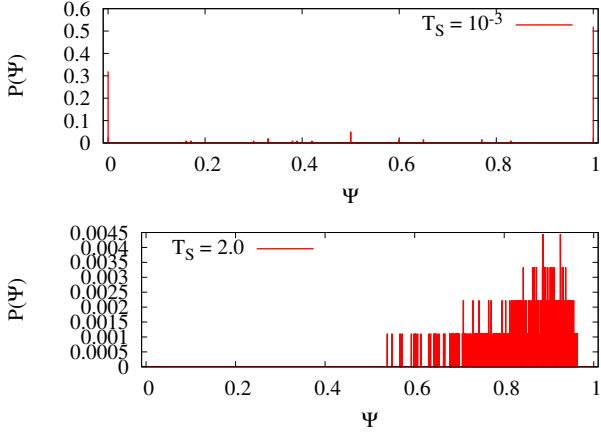}
\end{center}
\caption{The fitness distribution of mutated $\bm{J}$ with mutation rate $\mu = 0.1$,
which are generated from the evolved genotypes at $T_S=10^{-3}$ and
at $T_S = 2.0$ at $N=15$.
The distribution is obtained by 100 types of mutated $\bm{J}$.
}
\label{mutation_TS}
\end{figure}

To confirm this schematic picture
of the fitness landscape, we have numerically obtained the fitness distribution
of mutated $\bm{J}$s around the adapted $\bm{J}$. 
Here we have computed the fitness values $\Psi$ for $\bm{J}$s mutated from the adapted $J$
with the mutation rate $\mu=0.1$. 
In \Fref{mutation_TS}, we have plotted the distribution of fitness values at
$T_S=10^{-3}$ and $T_S=2.0$ for $N=15$ and $t=3$.
As shown, the fitness distribution
of mutated $J$s at low $T_S$ has two peaks at $\Psi = 1$ and $0$.
Hence some mutations are neutral, while others result in a sharp drop in the fitness
to its minimal values.
In contrast, the fitness of the mutated $\bm{J}$s at the intermediate
$T_S$ is broadly distribute around 0.8. This result
supports the schematic fitness landscape in \Fref{neutral_space}.

\begin{figure}
\begin{center}
\includegraphics[width=\figwidth]{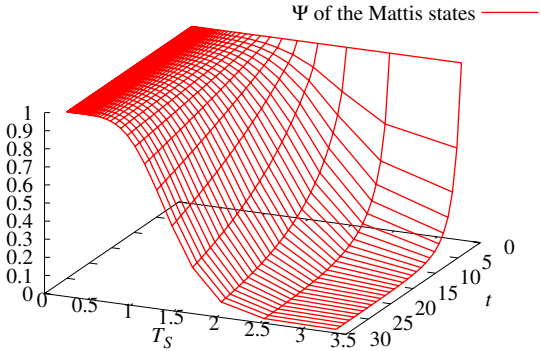}
\end{center}
\caption{$T_S$- and $t$-dependence of the fitness $\Psi(T_S,t)$ of the Mattis state
at $N=30$.}
\label{Mattis_vs_t}
\end{figure}

At the intermediate $T_S$, it has been shown that the LMS phase vanishes
at sufficiently large $t$ as seen in \Fref{N30}(c). This could be
understood by the $T_S$ and $t$-dependence of the fitness of
the Mattis state. The fitness of the Mattis state at $N=30$ is plotted in
\Fref{Mattis_vs_t}, which is obtained from \Eref{eqn:part-func} and
\Eref{eqn:Zf}.
The fitness of the Mattis state as well as the LMS one at the
intermediate temperature region decreases as $t$ is increased.
As shown in \Fref{N30}(a),
the fitness value decreases rapidly with $T_S$ as $t$ increases.
The drop with the increase of $T_S$
is more prominent for larger $t$, and thus the temperature interval to
support the LMS gets narrower with the increase of $t$, and is expected to
disappear for large $t$.

\section{Conclusions and Discussions}
\label{conclusions}
We have considered the evolution of a Hamiltonian
system to generate a specific configuration for target spins that
captures the basic features required to study the evolution.
In this study, we adopted a Markov process, which is given by temperature
$T_{J}$ and fitness $\Psi(\bm{J})$, for evolutional dynamics. By
performing numerical simulation, we found that a specific subset of
$\bm{J}$ with low energy and high fitness is evolved at an intermediate
$T_{S}$ and low $T_{J}$. From the statistical-mechanical viewpoint,
we focused on frustration and found that the interactions $\bm{J}$
that evolved at the intermediate $T_{S}$ are less frustrated. We
called these $\bm{J}$ the local Mattis states. In general, the less
frustrated $\bm{J}$ states are robust to mutation. Hence, the robustness
of evolving states to mutation is realized at intermediate temperatures
$T_{S}^{c1}\leq T_{S}\leq T_{S}^{c2}$. In other words, robustness
to thermal noise introduces mutational robustness; this has also been
recently discussed for gene regulation network models\cite{KK-PLoS}.
The relevance of thermal noise to robust evolution is thus demonstrated.

The mechanism by which the mutational robustness is achieved could
be understood by a statistical-mechanical argument. The $T_{S}$-dependence
of the fitness landscape was determined by explicitly calculating
the fitness $\Psi(\bm{J})$ for a typical case. It was found that
the correlation between fitness and mutational robustness is generated
at intermediate $T_{S}$, 
while it disappears at sufficiently low $T_{S}$. Hence, the mutationally
robust interactions $\bm{J}$ are obtained as a result of the selection
under a certain level of noise. The evolution of the mutational robustness
at the intermediate $T_{S}$ is confirmed by applying statistical-mechanical
theory.

We also found that for the interactions evolved at the intermediate
$T_{S}$, the relaxation to equilibrium progresses smoothly without
being stuck at metastable states. For protein folding, such an energy
landscape was proposed in terms of the consistency principle by Go \cite{Go}
and as a funnel-landscape by Onuchic et al. \cite{Onuchic}. The funnel
energy landscape has no ruggedness around the folded state, in contrast
to the spin-glass state. On the other hand, far from the folded state,
the ruggedness in the landscape remains; this is different from the
global attraction in the ferromagnetic or Mattis state. We note that
the landscape in a local Mattis state that is generated as a result
of evolution at $T_{S}^{c1}\leq T_{S}\leq T_{S}^{c2}$ is such a funnel
landscape, as there is no frustration around the target spins, whereas
frustration around non-target spins can result in ruggedness in the
landscape far from the target configuration. Indeed, such a smooth
and quick relaxation process is observed only for $T_{S}^{c1}\leq T_{S}\leq T_{S}^{c2}$,
whereas the relaxation is often stuck at metastable states for a system
evolved below $T_{S}^{c1}$. On the basis of this correspondence between
the funnel landscape and the local Mattis state, we expect that the
funnel landscape is characterized by a state with $\Phi_{1}=\Phi_{2}=1$.

Biological systems have evolved and functioned at a range of temperatures.
Recall that for a Hamiltonian evolved with $T_{S}<T_{S}^{c1}$, a
large number of time steps is required to reach a spin configuration
having the highest fitness, and the number of steps seems to increase
with the number of total spins. High fitness is not achieved for a
low-temperature region within a biologically acceptable time span,
i.e, within a single generation. Accordingly, adaptive evolution is
possible only when sufficient thermal noise is present; this corresponds
to $T_{S}>T_{S}^{c1}$.

We have found that the funnel-like landscape evolves in such biologically
relevant temperature regions. In fact, such a landscape is commonly
observed not only in protein folding but also in gene expression dynamics \cite{KK-PLoS,FangLi-Ouyang-Tang}
and in the morphogenesis of multicellular organisms \cite{KK-Asashima}.
Our result implies that a landscape that allows smooth relaxation
dynamics toward the target phenotype, such as the above-mentioned landscape,
is realized as a consequence of dynamics that are robust to thermal
noise as well as to mutation. We expect that this type of landscape
that results from robustness can be found in general in biological
evolution. It will not be restricted to Hamiltonian dynamics for protein
folding; rather, it will be generally applicable in
developmental dynamics. Indeed, recent studies on the evolution of
the gene regulation network also demonstrate that the funnel-type
dynamics evolve at the intermediate range of noise amplitude values
\cite{KK-PLoS,KK-chaos}. Our results may explain the ubiquity of
such funnel-like dynamics in evolved biological systems.

Although the frustration measure in the present paper is not directly
applied to the gene regulation network, the transition to the robust
developmental landscape is common. It will be interesting to study
the similarities and differences in the transitions in the spin Hamiltonian
model presented here and the dissipative gene expression dynamical
model.

It is also interesting that there is an optimal number of target spins
for achieving the local Mattis state with a funnel energy landscape
over a wide range of temperatures.
The number of non-target spins $N-t$ controls the redundancy of the system.
If $t$ is too small, the target configurations will be more easily
perturbed by non-target spins, and fitted states will be less robust
to the thermal noise.
On the other hand, if it is too large,
such $\bm{J}$ configurations with high fitness are too much limited in the
whole configurations,
and hence it will not be easily accessed or may be destabilized by
mutation. The existence of an appropriate number of redundant spins is
important to achieve
robustness. Indeed, in proteins, amino-acid residues that are responsible
for a function are limited and their number is typically smaller than
the number of other proteins.
Possible links between redundancy and the evolvability are pointed out
in \cite{Wagner-book},
but they still wait to be established quantitatively by theoretical analysis.
The present demonstration of the optimal fraction of
target elements for realizing robust functions will be important
in this context.

In this study, we observed transitions at $T_{S}^{c1}$ and $T_{S}^{c2}$.
The phase below $T_{S}^{c1}$ corresponds to the spin-glass phase
and that above $T_{S}^{c2}$ corresponds to the paramagnetic phase
in the context of statistical physics; in contrast, the local Mattis
phase between the two phases, which could correspond to the funnel
landscape, is a novel discovery in this study. Since a framework of
statistical physics of spin systems has been adopted in the present study, the
theoretical concepts developed therein, such as replica symmetry breaking,
may be applicable in the context developed here to understand this
transition. 
In particular, 
our model in this study is a variant
of spin Hamiltonian systems with two temperatures: one for spin and
the other for interactions $\bm{J}$. 
Theoretical analysis of such systems\cite{Sherrington,Dotsenko}
will be relevant for the analysis of the local Mattis state and mutational
robustness that we have discussed in this paper.

\begin{acknowledgments}
This study was partially supported by a Grant-in-Aid for Scientific Research (No.18079004) from MEXT
and JSPS Fellows (No.20$-$10778) from JSPS. 
\end{acknowledgments}

\newpage %Just because of unusual number of tables stacked at end
%\bibliography{SHK_biblio}% Produces the bibliography via BibTeX.
%\section*{References}

\end{document}